\documentclass[journal,comsoc]{IEEEtran}
\usepackage[T1]{fontenc}% optional T1 font encoding
\usepackage{algorithm} % algorithm package
\usepackage{algorithmic}
\usepackage{graphicx}
\usepackage{grffile}
\usepackage{verbatim}
\usepackage{bm}
\usepackage{amssymb}
\usepackage{cite}

\ifCLASSINFOpdf
\else
\fi
\usepackage{amsmath}
\usepackage[cmintegrals]{newtxmath}
\begin{document}
\title{Fine Timing and Frequency Synchronization for MIMO-OFDM:~An Extreme Learning Approach}

\author{Jun~Liu,~Kai~Mei,~Xiaochen~Zhang,~Des McLernon,~\IEEEmembership{Member, IEEE},~Dongtang Ma,~\IEEEmembership{Senior Member, IEEE},~Jibo~Wei,~\IEEEmembership{Member, IEEE}~and Syed Ali Raza Zaidi,~\IEEEmembership{Senior Member, IEEE}% <-this % stops a space
\thanks{This work was supported in part by the China Scholarship Council (CSC),~National Natural Science Foundation of China (NSFC) under Grant 61931020, 61372099 and 61601480.~(\emph{Corresponding author:~Dongtang Ma and Jun Liu.})}% <-this % stops a space
\thanks{Jun Liu,~Kai Mei,~Xiaochen Zhang,~Dongtang Ma,~and Jibo Wei are with the College of Electronic Science and Technology,~National University of Defense Technology,~Changsha~410073,~China~(E-mail:~\{liujun15,~meikai11,~zhangxiaochen14,~dongtangma,~wjbhw\}@nudt.edu.cn).}

\thanks{Des McLernon, Jun Liu and Syed Ali Raza Zaidi are with the School of Electronic and Electrical Engineering,~University of Leeds,~Leeds,~LS2 9JT, UK~(E-mail:~\{D.C.McLernon,~eljliu,~S.A.Zaidi\}@leeds.ac.uk).}}


\markboth{
	IEEE Transactions on Cognitive Communications and Networking,~Vol.~XX, No.~X, October~2021}%
{Shell \MakeLowercase{\textit{et al.}}: Bare Demo of IEEEtran.cls for IEEE Communications Society Journals}



\maketitle

\begin{abstract}
Multiple-input multiple-output orthogonal frequency-division multiplexing (MIMO-OFDM) is a key technology component in the evolution towards cognitive radio (CR) in next-generation communication in which the accuracy of timing and frequency synchronization significantly impacts the overall system performance. In this paper, we propose a novel scheme leveraging extreme learning machine (ELM) to achieve high-precision synchronization. Specifically, exploiting the preamble signals with synchronization offsets, two ELMs are incorporated into a traditional MIMO-OFDM system to estimate both the residual symbol timing offset (RSTO) and the residual carrier frequency offset (RCFO). The simulation results show that the performance of the proposed ELM-based synchronization scheme is superior to the traditional method under both additive white Gaussian noise (AWGN) and frequency selective fading channels. Furthermore, comparing with the existing machine learning based techniques, the proposed method shows outstanding performance without the requirement of perfect channel state information (CSI) and prohibitive computational complexity. Finally, the proposed method is robust in terms of the choice of channel parameters (e.g., number of paths) and also in terms of ``generalization ability'' from a machine learning standpoint.
\end{abstract}


% \fi
\IEEEpeerreviewmaketitle

\section{Introduction}
\IEEEPARstart{C}{ognitive} radio (CR) in fifth generation (5G) cellular communications technology for commercial use is currently being deployed in various countries. Meanwhile, research into sixth generation (6G) systems is already under way as it is designed to meet the needs of ultra-high capacity, reliability and low latency \cite{Rajatheva2020}. Among existing and future technologies, the key enabling technologies CR and multiple-input multiple-output orthogonal frequency-division multiplexing (MIMO-OFDM) will continue to play important roles that facilitate development and deployment. To improve spectral efficiency, CR enables nodes to explore and use underutilized licensed channels. And meanwhile, as an effective physical layer solution, the spectral efficiency (SE) of OFDM systems is superior to conventional single carrier systems and it can combat inter-symbol interference (ISI) through transforming a frequency-selective fading channel into many parallel flat-fading subchannels.

However, OFDM systems are highly sensitive to carrier frequency offset (CFO) which can destroy the important orthogonality between subcarriers and this results in the degradation of bit error rate (BER) performance. Therefore, the estimation of accurate CFO is crucial to OFDM systems. Meanwhile, symbol timing offset (STO) can result in ISI and a rotated phase whose value is proportional to the subcarrier index at the FFT output in an OFDM receiver. These issues severely degrade the efficiency of CR and the performance of communications systems \cite{balachander2021carrier}.

The traditional approach towards the estimation of both STO (also known as the timing synchronization) and CFO (also known as the frequency synchronization), involves sending a preamble at OFDM transmitters and processing the signals at the receivers. These signal processing techniques have been studied extensively, and many seminal articles have been published since the 1990s. P.H. Moose addressed the issue of receiver frequency synchronization by proposing an algorithm for a maximum likelihood estimate (MLE) of the CFO using the discrete Fourier transform (DFT) of a repeated symbol, and a lower bound for signal-to-noise (SNR) has been derived \cite{Moose1994}. In \cite{Schmidl1997}, a method for the rapid and robust frequency and timing synchronization for OFDM has been presented by Schmidl \textit{et al.} Then, an implementation of an MIMO-OFDM-based wireless local area network (WLAN) system was demonstrated by \cite{Zelst2004}, in which a simple MIMO extension of Schmidl’s algorithm \cite{Schmidl1997} proposed in \cite{Mody2001} was deployed in a practical system. In \cite{1703862} the authors address the problem of training design for a frequency-selective channel and also CFO estimation in single- and multiple-antenna systems under different energy-distribution constraints. In \cite{7582444} and \cite{7933261}, a new framework referred to as sparse blind CFO estimation for interleaved uplink orthogonal frequency-division multiple access (OFDMA) and the sparse recovery assisted CFO estimator for the uplink OFDMA were respectively proposed. Timing and frequency synchronization, as well as channel estimation can be carried out jointly to achieve better performance. \cite{7982654} presents a novel preamble-aided method for joint estimation of timing, CFO, and channel parameters for OFDM. \cite{8013752} considered the joint maximum likelihood estimator for the channel impulse response (CIR) and the CFO. In \cite{Nasir2016}, a comprehensive literature review and classification of the recent research progress in timing and carrier synchronization was presented.

But, errors will nearly always remain in the estimation of STO and CFO, which are also known as residual STO (RSTO) and residual CFO (RCFO). This is due to the effects of fading and thermal noise. The influence of STO errors on channel interpolation is analyzed in \cite{Chang2008}.~Even a small RCFO can result in amplitude and phase distortion and also inter-carrier interference (ICI) among subcarriers.~Traditionally, in order to mitigate the impact of RCFO, channel tracking methods are employed, and this is realized by inserting known pilots into specific subcarriers. However, this method reduces system SE. As regards RSTO, a compensation method for channel correction needs to be used. To reduce the sensitivity to synchronization errors, \cite{6963465} develops conditions for the selection of appropriate Zadoff-Chu sequences and then designs a training sequence and proposes joint signal detection, timing, and CFO estimation algorithms for OFDM downlink transmissions.

In recent years, challenges to traditional methods have emerged that use data-based approaches relying on machine learning \cite{daniels2009adaptation,sonal2016MachineLF, van2019deep, li2018carrier, he2020improved}. A popular scheme is machine learning-based end-to-end communications systems. Based on the idea of autoencoder, Dörner \textit{et al.} proposes a learning-based communications system, in which the task of synchronization is addressed through a neural network \cite{Dorner2018}. Similarly, in \cite{Wu2019}, a sampling time synchronization model using a convolutional neural network (CNN) for end-to-end communications systems is introduced. In \cite{Qing2020}, an extreme learning machine (ELM)-based frame synchronization method for a burst-mode communications system is proposed. The work in \cite{elwekeil2018deep} uses CNN to achieve adaptive modulation and coding in MIMO-OFDM systems with practical impairments, such as imperfect synchronization and channel estimation. ~Finally, \cite{Ninkovic2020}~investigates a deep neural network (DNN)-based solution for packet detection and CFO estimation.

Although the above-mentioned machine learning-based schemes achieve better performance or robustness than traditional methods, their shortcomings lead to serious difficulties in practical implementation. We summarize the challenges and deficiencies of these schemes as follows.
\begin{itemize}
	
	\item In previous machine learning related works \cite{daniels2009adaptation, sonal2016MachineLF, van2019deep, li2018carrier, he2020improved, Dorner2018}, the synchronization aspects are ignored and perfect synchronization is assumed. Since the trained parameters of the neural network depend significantly on the input data, when the test signal has STO and CFO, these methods would crash.
	
	\item  The mathematical theory of communication was exhaustively explored by Shannon in \cite{Shannon1948}, where the fundamental problem is described as ``reproducing at one point either exactly or approximately a message selected at another point''. But, autoencoder-based methods \cite{Dorner2018,Wu2019} present a ``chicken and egg'' problem. That is to say, you first need a reliable communications system to do the error back-propagation to actually train an end-to-end communications system for you. Therein lies the paradox.
	
	\item The schemes in \cite{Wu2019,elwekeil2018deep,Qing2020,Ninkovic2020} sink into another kind of ``chicken and egg'' dilemma. Specifically, in the training stage, they  require the accurate probability density function (PDF) of a channel to generate labeled data with exact timing location and CFO. Unfortunately, it is impossible to priorly acquire the PDF of a fading channel in real scenarios. Besides, to train these schemes under a real channel, the target output, exact timing location and CFO are impossible to acquire.
	
	\item The common disadvantage of most of the current learning-based techniques lies in the computational complexity because they are based on a DNN. DNNs usually have deep hidden layers, which requires prohibitive computational complexity.
\end{itemize}

Motivated by the challenges mentioned above, in this paper we first propose a robust ELM-based fine timing and frequency synchronization scheme to deal with the challenges above. Unlike previous learning schemes, where a reliable feedback link between the transmitter and receiver or an accurate channel PDF is necessary, we deploy an ELM at the receiver and we generate training data by letting the preamble be corrupted by RSTO and RCFO without the effects of multipath fading channel and thermal noise. There are two main reasons for this. The first aims to make ELM learn the relationship between the corrupted preamble and its corresponding RSTO and RCFO. The second reason is to avoid the need for the channel PDF. Specifically, we combine ELMs to a typical MIMO-OFDM system instead of using an end-to-end autoencoder-based model so that the training can be carried out entirely at the receiver. Then, the estimation of both STO and CFO relay on that the preamble signal is perfectly known for the receiver and it consists two identical parts. Therefore, we let the ELMs to learn the sequences of the original preamble and the corrupted preambles with RSTO and RCFO instead of sequences of the preambles corrupted by a fading channel with a given PDF. These two strategies prevent our proposed schemes sink into the ``chicken and egg'' dilemma. In addition, an ELM only has one hidden layer so its computational complexity of training and prediction is significantly lower than a DNN. The main contributions of this paper are summarized as follows. 
\begin{itemize}
	\item For MIMO-OFDM, we incorporate ELM with a traditional STO estimator. In the proposed scheme, coarse synchronization is carried out by using autocorrelation-based algorithm. Then, the fine timing synchronization can be achieved by ELM without the need for any prior information about the channel. 
	
	\item We first propose a robust ELM-based scheme to realize RCFO estimation without the need for additional prior information about the channel, where the ELM can learn the mapping relationship between the preamble corrupted by both RSTO and RCFO.
	
	\item We give the comparison of the complexity between the proposed ELM-based models and the traditional algorithm and DNN-based methods. Then, the performance analysis of the proposed learning scheme in different cases is provided. Specifically, computer simulation results show that the proposed scheme is superior to traditional STO and CFO estimation methods in terms of mean squared error (MSE). In addition, extensive simulation results and comparisons have demonstrated the robustness and (machine learning) generalization ability of the proposed scheme.
\end{itemize}

%So, how do we deal with the dilemma? Most of the existing machine learning-based physical techniques require labeled data with perfect CSI in the training stage. The question raised in these studies is that it is impossible to acquire (estimate) the perfect CSI in real channel scenarios due to the existence of thermal noise. In other words, which came first: the labeled data or the prefect CSI? The essential reason for this dilemma is that CSI is random and therefore it cannot be estimated or predicted perfectly. Besides, the computational cost to learn the relationship between random variables and the target output is extremely high. Nevertheless, can we deal with this causality dilemma?%

The remainder of this paper is organized as follows. The signal model of the MIMO-OFDM system and traditional timing and frequency synchronization for MIMO-OFDM are presented in Sections \ref{S2} and \ref{S3}, respectively. In Section \ref{S4}, we propose a scheme that incorporates ELM into the traditional MIMO-OFDM system, in which ELM is used to estimate RSTO and RCFO. Then, numerical results and analysis for evaluating the performance of the proposed scheme are provided in Section \ref{S5}, which is followed by conclusions in Section \ref{S6}.

\textit{Notations:}~The notations adopted in the paper are as follows. We use boldface lowercase $\bf{x}$ and capital letters $\bf{X}$ to denote column vectors and matrices, respectively. Superscripts $^{-1}$,~$^*$,~$^T$,~$^H$ and~$^\dagger$ stand for inverse, conjugate, transpose,~Hermitian transpose and Moore-Penrose pseudoinverse, respectively. In addition,~$\otimes$,~$  \odot  $,~$\circledast$, ${\rm E}\left\{\cdot\right\}$,~$ \lfloor \cdot \rfloor $~and $j=\sqrt{-1}$ denote respectively the Kronecker product,~Hadamard product,~cyclic convolution, the expectation operation, floor function and the imaginary unit.~Note that $ \angle \left(  \cdot  \right) $~returns the phase angle of a complex number. Finally, $ {\rm{repmat}}~({\bf{A}},m,n) $ returns an array containing $ m $ and $ n $ copies of $ {\bf{A}} $ in the column and row dimensions, respectively.

\section{MIMO-OFDM Signal Model}
\label{S2}
Let us now consider a MIMO-OFDM system with $ N_t $ transmit (TX) and $ N_r $ receive (RX) antennas, which is usually denoted as a~$ N_t \times N_r $~system. Without loss of generality, we consider the frequency-domain MIMO-OFDM signal model, which is directly given as~\cite{Zelst2004}
\begin{equation}
{\bf{\tilde x}}\left( a \right) = {\bf{\tilde H\tilde s}}\left( a \right) + {\bf{\tilde n}}\left( a \right)
\end{equation}
where an $ N_rN_c $-dimensional complex vector $ {\bf{\tilde x}}\left( a \right) $ represents the frequency-domain received signal, $ {\bf{\tilde s}}\left( a \right) = {\left[ {{\bf{s}}{{\left( {0,a} \right)}^T}, \cdots, {\bf{s}}{{\left( {{N_c} - 1,a} \right)}^T}} \right]^T} \in {^{{N_tN_c} \times {1}}} $ and $ {\bf{s}}\left( {k,a} \right) $ represents an $ N_t $-dimensional complex vector transmitted on the $ k $th subcarrier of the $ a $th MIMO-OFDM symbol with $ {S_p}\left( {k,a} \right) $ as its $ p $th element, i.e., transmitted on the the $ p $th TX antenna. $ {\bf{\tilde n}}\left( a \right) $ represents the frequency-domain noise vector, with i.i.d. zero-mean, complex Gaussian elements with variance $ 0.5\sigma _n^2 $ per dimension, and the channel frequency response is represented as a block diagonal matrix  $ {{\bf{\tilde H}}} $  as follows:
\begin{equation}
{\bf{\tilde H}} = \left[ {\begin{array}{*{20}{c}}
	{{\bf{H}}\left( 0 \right)}&{}&0\\
	{}& \ddots &{}\\
	0&{}&{{\bf{H}}\left( {{N_c} - 1} \right)}
	\end{array}} \right].
\end{equation}
Now, $ {\bf{H}}\left( k \right) \in \mathbb{C}{^{{N_r} \times {N_t}}} $ represents the $ {{N_t} \times {N_r}} $ MIMO channel for the $ k $th subcarrier and can be shown to be
\begin{equation}
{\bf{H}}\left( k \right) = \sum\limits_{l = 0}^{L - 1} {{\bf{G}}\left( l \right)\exp \left( { - j2\pi \frac{{kl}}{{{N_c}}}} \right)}
\end{equation}
where the $ l $th path of MIMO CIR matrix $ {\bf{G}}\left( l \right) \in \mathbb{C}{^{{N_r} \times {N_t}}} $ and its $ \left( {q,p} \right) $th element is $ {{g_{q,p}}\left( l \right)} $.
We assume that these taps are independent, zero-mean, complex Gaussian random variables with variance $ 0.5{P_l} $ per dimension.~The ensemble $ {P_l},~l = \left\{ {0, \cdots ,L - 1} \right\} $ is called the power delay profile (PDP) and its total power is assumed to be normalized to $ \sigma _c^2 = 1 $.~For each $ k $th subcarrier, the signal model can be written in its flat-fading form as
\begin{equation}
{\bf{x}}\left( {k,a} \right) = {\bf{H}}\left( k \right){\bf{s}}\left( {k,a} \right) + {\bf{n}}\left( {k,a} \right).
\end{equation}

\section{Traditional Timing and Frequency Synchronization for MIMO-OFDM}
We consider the traditional preamble pattern \cite{Schmidl1997} and synchronization method \cite{Zelst2004} in this section.
\label{S3}
\begin{figure}[t]
	\centering
	\includegraphics[width=3.5in]{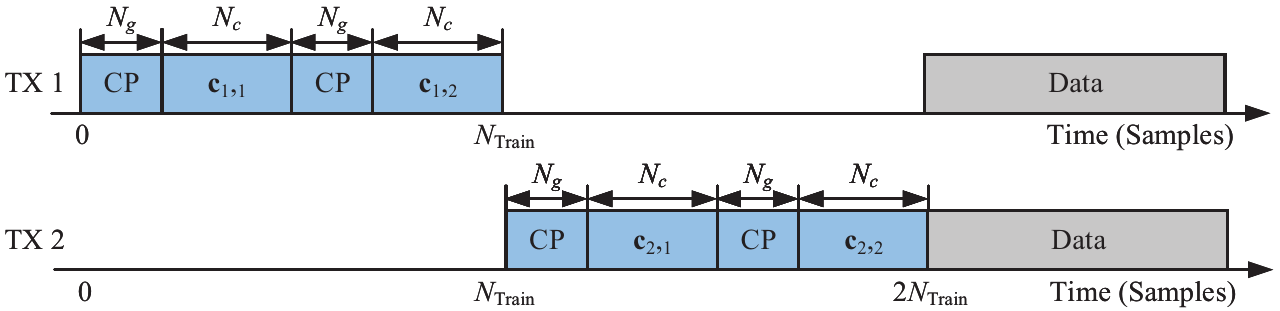}
	\caption{Structure of a time orthogonal preamble for a $ 2\times2 $~MIMO-OFDM system.}
	\label{PreambleStructure}
\end{figure}
As shown in Fig. \ref{PreambleStructure}, in order to estimate the subchannels between the different TX and RX antennas, a time orthogonal preamble is chosen. The length of the preamble for all the TX antennas is $ {N_{{\rm{train}}}} = 2\left( {{N_g} + {N_c}} \right) $, where  $ N_g $ and $ N_c $ denote the length of the cyclic prefix (CP) and one OFDM symbol, respectively. $ {\bf{c}}_{p,1} $ and $ {\bf{c}}_{p,2} $ are different pseudo-noise (PN) sequences transmitted by the $ p $th TX. 

The first part of the preamble $ \bm{c}_{p,1} $ comprises two identical halves in the time domain, which are used for symbol timing and fractional CFO estimation. This kind of time-domain identical structure can be obtained by transmitting a PN sequence only on the even frequencies while zeros are placed on the odd frequencies. The second part of the preamble $ \bm{c}_{p,2} $ contains a PN sequence on its odd frequencies to measure these subchannels and another PN sequence on the even frequencies to help determine the frequency offset.
\subsection{Timing Synchronization}
Before the estimation of the CFO is conducted, the STO~$\left( \tau  \right)$~needs to be estimated.~The method for timing synchronization is given by \cite{Schmidl1997,Zelst2004}
\begin{equation}
\hat \tau  = \mathop {{\rm{argmax}}}\limits_d \frac{1}{N_g}\sum\limits_{m = 0}^{N_g - 1} {\left[ {\frac{{\sum\limits_{p = 1}^{{N_t}} {{{\left| {\Lambda \left( {{d_p} + m} \right)} \right|}^2}} }}{{\sum\limits_{p = 1}^{{N_t}} {P{{\left( {{d_p} + m} \right)}^2}} }}} \right]} ,
\end{equation}
where ${d_p} = d - \left( {{N_t} - p} \right){N_{{\rm{train}}}}$ and $ d $ are discrete variables.~$\Lambda \left( d \right)$ is the complex autocorrelation of the first part of preamble~$ \bf{c_1} $,~and is given by
\begin{equation}
\Lambda \left( d \right) = \sum\limits_{i = d - \left( {{N_c}/2 - 1} \right)}^d {\sum\limits_{q = 1}^{{N_r}} {r_q^*\left( {i - {N_c}/2} \right){r_q}\left( i \right)} }
\end{equation}
with $ r_q(i) $ the $ i $th sample of the received signal on the $ q $th antenna. The received energy for the second half-symbol of $ \bf{c_1} $,~$P\left( d \right)$,~is defined by
\begin{equation}
P\left( d \right) = \sum\limits_{i = d - \left( {{N_c}/2 - 1} \right)}^d {\sum\limits_{q = 1}^{{N_r}} {r_q^*\left( i \right){r_q}\left( i \right)} } .
\end{equation}
Note that $ d $~is a time index corresponding to the first sample in a window of $ N_c $ samples.
\subsection{Frequency Synchronization}
In this subsection, the CFO estimation method is based on \cite{Schmidl1997}. We define a normalized CFO,~$ \varepsilon $,~as a ratio of the CFO $ f_{\rm{offset}} $ to subcarrier  spacing $ \Delta f $, given as $ \varepsilon = f_{\rm{offset}}/ \Delta f$. Let $ \varepsilon_i $ and $ \varepsilon_f $ denote the integer part and fractional part of $ \varepsilon $, respectively, and therefore $ \varepsilon = \varepsilon_i + \varepsilon_f $, where $ \varepsilon_i = \lfloor \varepsilon \rfloor $.~If~$ \left| \varepsilon  \right| \le 1 $, the CFO can be estimated directly as
\begin{equation}
\hat \varepsilon {\rm{ = }}\frac{{\hat \theta }}{\pi }{\rm{ = }}\frac{{\angle \left[ {\sum\limits_{p = 1}^{{N_t}} {\Lambda \left( {{{\hat \tau }_p}} \right)} } \right]}}{\pi },
\label{Estimation CFO}
\end{equation}
where ${{\hat \tau }_p} = \hat \tau  - \left( {{N_t} - p} \right){N_{{\rm{train}}}}$ and $ \hat{\theta} $ denotes the phase of the summation of the complex correlations of the preambles originating from the different transmitters. When $ \left| \varepsilon  \right| > 1 $, the PN sequence  on the  even frequencies of $ \bf{c}_2 $ will be needed and the CFO can be given by
\begin{equation}
\varepsilon {\rm{ = }}\frac{\theta }{\pi } + 2g,
\end{equation}
where $ g $ is an integer. By partially correcting the frequency offset, adjacent carrier interference can be avoided, and then the remaining offset of $ 2g $ can be found. In order to estimate $ g $, the received preamble at the $ q $th RX antenna from the $ p $th TX antenna, corresponding to $ {\bf{c}}_{p,1} $ and $ {\bf{c}}_{p,2} $ needs to be frequency compensated by  $ \hat{\theta} $ at first and then transformed into the frequency domain as $ {\bf{x}}_{q,p,1} $ and $ {\bf{x}}_{q,p,2} $, respectively. Then, $ g $ can be estimated by the difference correlation as follows:
\begin{equation}
\resizebox{.99\hsize}{!}{$ \hat g = \mathop {\arg \max }\limits_g \frac{{\sum\limits_{p = 1}^{{N_t}} {\sum\limits_{q = 1}^{{N_r}} {{{\left| {\sum\limits_{k \in {{\rm X}_{{\rm{Even}}}}} {X_{q,p,1}^*\left[ {k + 2g} \right]v_p^*\left[ k \right]{X_{q,p,2}}\left[ {k + 2g} \right]} } \right|}^2}} } }}{{2\sum\limits_{p = 1}^{{N_t}} {\sum\limits_{q = 1}^{{N_r}} {{{\left( {\sum\limits_{k \in {{\rm X}_{{\rm{Even}}}}} {{{\left| {{X_{q,p,2}}\left[ k \right]} \right|}^2}} } \right)}^2}} } }}, $}
\end{equation}
where $X_{\rm{Even}}$ represents the subset of even frequency indices and $ {v_p}\left[ k \right] = \sqrt 2 {c_{p,2}}\left[ k \right]/{c_{p,1}}\left[ k \right],~k \in X_{\rm{Even}} $. Finally, the estimate can be written as 
\begin{equation}
\hat \varepsilon  = \frac{{\hat \theta }}{\pi } + 2\hat g.
\end{equation}
\section{ELM-Based RSTO and RCFO Estimation}
\label{S4}
Due to the fading channel and thermal noise, small but significant RSTO and RCFO will always exist to degrade the performance of MIMO-OFDM systems. In order to perform synchronization more accurately, the methods of ELM-based RSTO and RCFO estimations will be introduced in this section.

\subsection{ELM-Based RSTO Estimation}

\begin{figure}[t]
	\centering
	\includegraphics[width=3.6in]{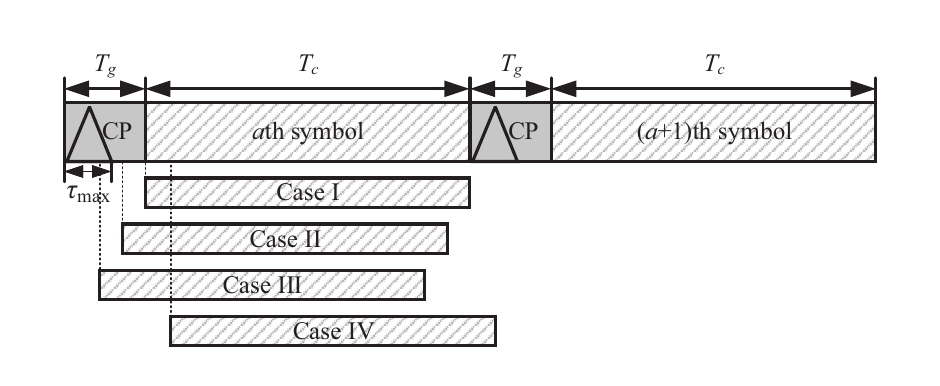}
	\caption{Four different cases of an OFDM symbol starting point subject to STO.}
	\label{EffectsOfSTO}
\end{figure}

Inspired by the idea that a neural network (NN) can learn from appropriate data, we try to further exploit (by a NN) the implicit information inside the preamble to estimate RSTO and RCFO. Compared with a DNN, an ELM only has single hidden layer and thus it has lower computational complexity, but it still has excellent performance \cite{Liu2019}. Specifically, the ELM model is chosen by comparing it with other machine learning models that are applied for communications signal processing in recent literature, mainly including DNN and CNN. Both DNN and CNN are typical deep learning technique, while ELM has a very simple network structure and is the
most promising technique to reduce the required training data and computational complexity, which contributes to reduce latency. In that regard, we choose to employ ELM models in this paper. Specifically and most importantly, we expect that the relationships between the corrupted preamble signal and synchronization offset can be ``learnt'' by the ELM. Therefore, it is necessary to first explain the effect of STO.

Depending on the location of the estimated starting point of an OFDM symbol, the effect of STO can vary. Fig. \ref{EffectsOfSTO} shows four different cases of timing offset, in which the estimated starting point is perfectly accurate ({\bf{Case I}}), a little early ({\bf{Case II}}), too early ({\bf{Case III}}), or a little late compared to exact timing ({\bf{Case IV}}). $ T_c $, $ T_g $ and $ \tau_{\rm{max}} $ represent the duration of the OFDM symbol, the CP and the maximum excess delay, respectively \cite{cho2010mimo}.

In {\bf{Case II}}, the channel response to the $ (a-1) $th OFDM symbol does not overlap with the $ a $th OFDM symbol and so does not incur any ISI from the previous symbol. In this case, the received signal in the frequency domain is obtained by taking the FFT of the time domain received samples:
\begin{equation}
\label{CaseII}
{\bf{x}}\left( {k,a} \right) = {\bf{H}}\left( k \right){\bf{s}}\left( {k,a} \right){e^{j2\pi k\tau /N}} + {\bf{n}}\left( {k,a} \right),
\end{equation}
where $ \tau $ denotes the STO. Equation (\ref{CaseII}) implies that the orthogonality among subcarrier frequency components can be completely preserved. However, there exists a phase offset that is proportional to the STO $ \tau $ and subcarrier index $ k $, forcing the signal constellation to be rotated around the origin in the complex plane. 

In {\bf{Case III}} and {\bf{Case IV}}, the orthogonality among subcarrier components is destroyed by the ISI from the previous and the succeeding OFDM symbols, respectively. In addition, ICI will occur. A quantitative analysis of the ISI and ICI resulting from STO has been exhaustively studied in \cite{cho2010mimo}.

In order to avoid the occurrence of the {\bf{Case IV}} in Fig. \ref{EffectsOfSTO},  the timing point will be set $ N_g/4 $ points ahead of the estimated value from the traditional estimator in the case of a fading channel,\footnote{In order to avoid ISI, the FFT window start position has to be put in advance of the estimated point obtained by the coarse STO estimation algorithm \cite{Chang2008}.} whereas it still cannot entirely eliminate the RSTO.

Therefore, a natural idea is that using ELM to learn the relationship between received preamble with ISI, ICI  and RSTO $ {\tau _{\rm{R}}} $, where $ {\tau _{\rm{R}}}{\rm{ = }}\tau  - \hat \tau  $. Compared with DNN, ELM is considered as a general form of single layer feedforward neural networks, where the input weights and hidden layer biases of ELM are randomly generated. In other words, hidden layer outputs are always known. Hence, this structure allows the analytical calculation of the output weights during the training phase by means of least square solutions. As a result, ELM has a competitive advantage in terms of computational complexity.

\begin{figure}[t]
	\centering
	\includegraphics[width=3.5in]{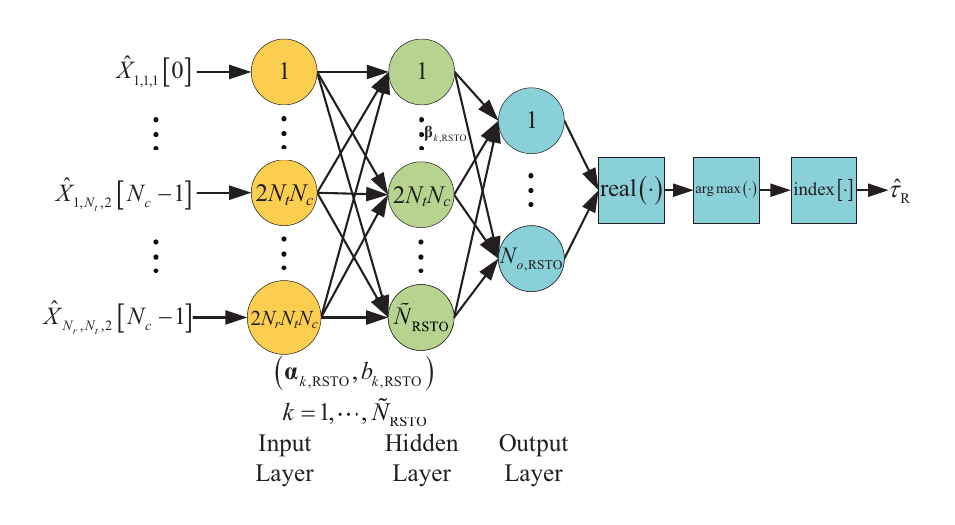}
	\caption{The structure of an ELM-based RSTO estimator.}
	\label{ELM-basedRSTOestimator}
\end{figure}

Under most situations, the starting point of an OFDM symbol will be one of the above-mentioned four cases. In other words, the probability of $\left|{\tau _{\rm{R}}}\right|>N_g$ is very low, and this fact, is very important to the generation of the training set for ELM-based RSTO estimator. Specifically, the range of the target values (RSTO) should be set properly but not too small or large. Otherwise, an inappropriate training set leads to poor performance of the ELM model. The details about the ELM-based model for RSTO estimation and the generation of training set is given as follows.

The structure of a complex ELM-based RSTO estimator is illustrated in Fig. \ref{ELM-basedRSTOestimator}.~The ELM-based RSTO estimator has $ 2N_rN_tN_c $ input neurons, $ \tilde N_{\rm{RSTO}} $ hidden neurons and $ N_{o,{\rm{RSTO}}} $ output neurons. The input, output and weights of ELM can be fully complex.~$ {\hat X_{q,p,i}}\left[ k \right] $ is the input in the prediction stage, which denotes the equalized frequency domain received signal at the $ q $th RX antenna from the $ p $th TX antenna corresponding to $ {{\bf{c}}_{p,i}} $. The $ {\rm{real}}\left(  \cdot  \right) $ block returns the real part of the elements of the complex array and the $ \arg \max \left(  \cdot  \right) $ block returns the indices of the maximum values.~The principle of the ELM-based RSTO estimator can be divided into two main stages: training and prediction stages.

%The data formats of the input in training and prediction stages are given in \eqref{FFTandRemoveCP}-\eqref{Cp=repmat} and \eqref{X}, respectively.%

\subsubsection{Training Stage}In this stage, the training set $ {\bf{N}_{\rm{RSTO}}} = \left\{ {\left( {{{{\bf{\tilde X}}}_n},{{\bf{O}}_{n,{\rm{RSTO}}}}} \right)|n = 1, \cdots ,N_{\rm{RSTO}}} \right\} $ is first generated. The $ n $th input data of training set $ {{\bf{\tilde X}}_n} \in {{\mathbb{C}}^{2{N_r}{N_t}{N_c} \times 1}}$ is given as 
\begin{equation}
	{{{\bf{\tilde X}}}_n} = {\mathop{\rm FFT}\nolimits} \left( {{\mathop{\rm Remove}\nolimits} {\mathop{\rm CP}\nolimits} \left( {{{\bf{X}}_n}} \right)} \right),
	\label{FFTandRemoveCP}
\end{equation}
where ${{\bf{\tilde X}}_n}$ denotes the combination  vector of the preamble signal in the frequency domain by taking the FFT of the time domain received samples with corresponding RSTO. In Equation (\ref{FFTandRemoveCP}), the pseudo-function ``RemoveCP'' and ``FFT'' represent respectively removing all the CPs and taking the $ N_c $-point fast Fourier transform. Specifically, the time domain received samples with RSTO ${\tau _{{\rm{R}},n}}$ can be expressed as 
\begin{equation}
	{{\bf{X}}_n} = {\bf{\tilde c}}\left[ k \right] \otimes \delta \left[ {k - {\tau _{{\rm{R}},n}}} \right],
	\label{DelayPreamble}
\end{equation}
where 
\begin{equation}
	{\bf{\tilde c}} = {\left[ {{{{\bf{\tilde c}}}_1}^T, \cdots ,{{{\bf{\tilde c}}}_p}^T, \cdots ,{{{\bf{\tilde c}}}_{{N_t}}}^T} \right]^T},
\end{equation}
and
\begin{equation}
	{{\bf{\tilde c}}_p} = {\rm{repmat}}\left( {{{\left[ {{\rm{CP}}_{{{\bf{c}}_{p,1}}}^T,{{\bf{c}}_{p,1}}^T,{\rm{CP}}_{{{\bf{c}}_{p,2}}}^T,{{\bf{c}}_{p,2}}^T} \right]}^T},{N_r},1} \right)
	\label{Cp=repmat}.
\end{equation}
Note that, in Equation (\ref{DelayPreamble}),~$ \delta \left[ {k - {\tau _{{\rm{R}},n}}} \right] $ denotes a delayed Kronecker delta function. The absent elements of $ {{\bf{\tilde c}}} $ will be filled by zero padding.

The $n$th target output, $ {{\bf{O}}_{n,{\rm{RSTO}}}} $ is a one-hot vector including encoded information of corresponding RSTO $ {\tau _{{\rm{R,}}n}} $. Now, for example, $ {\left[ {1,0,\cdots,0} \right]^T} $ represents $ {\tau _{\rm{R}}} =  - {N_g} $, $ {\left[ {0,\cdots,0,1} \right]^T} $ represents $ {\tau _{\rm{R}}} =   {N_g} $ and
\begin{equation}
{\tau _{{\rm{R}},n}} = {\rm{index}}\left[ {\mathop {\arg \max }\limits_{1 \le i \le 2{N_g} + 1} \left( {{o_{n,{\rm{RSTO}},i}}} \right)} \right],
\end{equation}
where $ {{\bf{O}}_{n,{\rm{RSTO}}}} = {\left[ {{o_{n,{\rm{RSTO}},1}},{o_{n,{\rm{RSTO}},2}}, \cdots ,{o_{n,{\rm{RSTO}},{2{N_g} + 1}}}} \right]^T} $.
Here~``\rm{index}'' represents an index array including different values of RSTO. In this paper, the size of the training set for the ELM-based RSTO estimator is $2N_g+1$ and $ {\rm{index}} = [ - {N_g}, \cdots ,{N_g}] $.

Then, the training set can be used for the determination of the weights and biases of the ELM. In {\bf{step 1}} of {\bf{Algorithm \ref{alg:1}}} (see later), the complex input weight $ {\bm{\alpha}_{k,{\rm{RSTO}}}} $ and complex bias $ b_{k,{\rm{RSTO}}} $ (see also Fig.\ref{ELM-basedRSTOestimator}) are generated from the uniform  distribution $ U\left( { - 0.1,0.1} \right) $, where ${\bm{\alpha}}_{k,{\rm{RSTO}}}\in{\mathbb{C}}^{2N_rN_tN_c  \times 1}$ is the input weight vector connecting input neurons to the $k$th hidden neuron and $ {\bm{\alpha }_{\rm{RSTO}}} = \left[ {{{\bm{\alpha }}_{1,{\rm{RSTO}}}},\cdots, {{\bm{\alpha }}_{k,{\rm{RSTO}}}}, \cdots ,{{\bm{\alpha }}_{{\tilde N_{\rm{RSTO}}},{\rm{RSTO}}}}} \right] $ and $ {\bf{b}_{\rm{RSTO}}} = \left[ {{b_{1,{\rm{RSTO}}}}, \cdots ,{b_{k,{\rm{RSTO}}}}, \cdots ,{b_{\tilde N_{\rm{RSTO}},{\rm{RSTO}}}}} \right] $. Once the input weights and biases are chosen, the output of the hidden layer can be given by

\begin{equation}
{{\bf{D}}_{{\rm{Training,RSTO}}}} = {g_c}\left( {{{\bm{\alpha }}^T_{{\rm{RSTO}}}}{\bm{{\rm \tilde X}}} + {\bf{b}}_{\rm{RSTO}}} \right)
\end{equation}
where $ {\bf{\tilde X}} = \left[ {\begin{array}{*{20}{c}}
	{{{\bf{\tilde X}}_1}}&{{{\bf{\tilde X}}_2}}& \cdots &{{{\bf{\tilde X}}_N}}
	\end{array}} \right] \in {{\mathbb{C}}^{2{N_r}{N_t}{N_c} \times N }}  $.

We expect that the output of the ELM could be close to the target output $ \bf{O}_{\rm{RSTO}} $, so
\begin{equation}
{\bm{\beta }}_{\rm{RSTO}}{{\bf{D}}_{{\rm{Training,{\rm{RSTO}}}}}} = {\bf{O}}_{\rm{RSTO}}.
\end{equation}
Generally,~$ {\bm{\beta }}_{\rm{RSTO}} = \left[ {{{\bm{\beta }}_{1,{\rm{RSTO}}}}, \cdots ,{{\bm{\beta }}_{k,{\rm{RSTO}}}}, \cdots ,{{\bm{\beta }}_{{\tilde N_{\rm{RSTO}}},{\rm{RSTO}}}}} \right] \in {{\mathbb{C}}^{{N_{o,{\rm{RSTO}}}} \times \tilde N_{\rm{RSTO}}}} $ and $ {{\bm{\beta }}_{k,{\rm{RSTO}}}} = {\left[ {{\beta _{k,{\rm{RSTO}},1}},{\beta _{k,{\rm{RSTO}},2}}, \cdots ,{\beta _{k,{\rm{RSTO}},{N_{o,{\rm{RSTO}}}}}}} \right]^T} \in {{\mathbb{C}}^{{N_{o,{\rm{RSTO}}}} \times 1}} $, where $ {{\bm{\beta }}_{k,{\rm{RSTO}}}} $ denotes the output weight vector connecting the $ k $th hidden neuron and the output neurons and $ N_{o,{\rm{RSTO}}} $ denotes the number of output neurons. For the ELM-based RSTO estimator, $ N_{o,{\rm{RSTO}}}=2N_g+1$. Under the criterion of minimizing the squared errors, the least squares (LS) solution is given by
\begin{equation}
	\begin{aligned}
		{\bm{\hat \beta }}_{\rm{RSTO}} &= \mathop {\rm{argmin} }\limits_{{\bm{\beta }}_{\rm{RSTO}}} \left\| {{\bm{\beta }}_{\rm{RSTO}}{{\bf{D}}_{{\rm{Training,RSTO}}}} - {\bf{O}}_{\rm{RSTO}}} \right\|\\
		 &= {\bf{O_{{\rm{RSTO}}}D}}_{{\rm{Training,RSTO}}}^\dag.
	\end{aligned}
\end{equation}
The training algorithm for an ELM-based RSTO estimator can be summarized as shown in \textbf{Algorithm \ref{alg:1}}. 
\begin{algorithm}[t]
	\caption{The Training Algorithm for an ELM-based RSTO Estimator}
	\label{alg:1}
	\begin{algorithmic}
		\STATE We are given a training set ${\bf{N}}_{\rm{RSTO}} = \left\{ {\left( {{{{\bf{\tilde X}}}_n},{{\bf{O}}_{n,{\rm{RSTO}}}}} \right)|n = 1, \cdots ,N_{\rm{RSTO}}} \right\}$, complex activation function $g_c\left( \cdot \right)$, and hidden neuron number $\tilde N_{\rm{RSTO}}$. ${\bf{\tilde X}}_n \in {{\mathbb{C}}^{2{N_r}{N_t}{N_c} \times 1}}$,~$ {{\bf{O}}_{n,{\rm{RSTO}}}} $ is a one-hot vector and these two correspond to the input and desired output of the ELM-based RSTO estimator, respectively.
		\STATE {\bf{Step 1:}} Randomly choose the values of complex input weight ${\bm{\alpha}}_{k,{\rm{RSTO}}}$ and the complex bias $b_{k,{\rm{RSTO}}}$, $k=1,\cdots,\tilde N_{\rm{RSTO}}$.
		\STATE {\bf{Step 2:}} Calculate the complex hidden layer output matrix ${\bf{D_{\rm{Training,RSTO}}}}$.
		\STATE {\bf{Step 3:}} Calculate the complex output weight ${\bm{\beta }}_{\rm{RSTO}}$ using $\bm{\hat \beta}_{\rm{RSTO}}={\bf{O}}_{\rm{RSTO}}{\bf{D}}^\dag_{\rm{Training,RSTO}} $, where $ {\bf{O}}_{\rm{RSTO}} \in {{\mathbb{C}}^{{N_{o,{\rm{RSTO}}}} \times N_{\rm{RSTO}}}} $.
	\end{algorithmic}
\end{algorithm}
\subsubsection{Prediction (Estimation) Stage}
For an ELM-based RSTO estimator, we assume that LS channel estimation is used or the perfect CSI is known. Thus the equalized preamble can be given by\footnote{A minimum mean-square error (MMSE) channel estimator cannot be deployed before the STO is estimated because the STO can degrade the performance of the MMSE channel estimator \cite{VanDeBeek1995}. However, the simulation results in Section \ref{S5} still include the ELM-based STO estimator with MMSE channel estimation. The introduction of the MMSE channel estimator is given in the next subsection.}
\begin{equation}
{\hat X_{q,p,i}}\left[ k \right] = {X_{q,p,i}}\left[ k \right]/{{\bf{H}}_{q,p,i}}\left[ k \right].
\end{equation}

The output of hidden layer can be calculated as
\begin{equation}
{{\bf{D}}_{{\rm{Prediction,RSTO}}}} = {g_c}\left( {{{\bm{\alpha }}^T_{\rm{RSTO}}}{{\bf{\hat x}}} + {\bf{b}}_{\rm{RSTO}}} \right)
\label{Prediction}
\end{equation}
where
\begin{equation}
{\bf{\hat x}} = {\left[ {{\bf{\hat x}}_{_{1,1,1}}^T,{\bf{\hat x}}_{_{1,1,2}}^T, \cdots ,{\bf{\hat x}}_{_{{N_t},{N_r} - 1,1}}^T,{\bf{\hat x}}_{_{{N_t},{N_r} - 1,2}}^T,{\bf{\hat x}}_{_{{N_t},{N_r},1}}^T,{\bf{\hat x}}_{_{{N_t},{N_r},2}}^T} \right]^T}.
\label{X}
\end{equation}
Note that the input, output and all the weights and biases of ELM in this paper are complex values but RSTO and RCFO are real values. Therefore, the operator ${\rm{real}}\left(  \cdot  \right)$ following the output of ELM is necessary. By expressing $ {\rm{real}}\left( {\bm{\hat \beta}_{\rm{RSTO}} {{\bf{D}}_{{\rm{Prediction,RSTO}}}}} \right) $ as $ {\rm{real}}\left( {\bm{\hat \beta}_{\rm{RSTO}} {{\bf{D}}_{{\rm{Prediction,RSTO}}}}} \right) = {\left[ {{{\hat o}_{1,{\rm{RSTO}}}},{{\hat o}_{2,{\rm{RSTO}}}}, \cdots ,{{\hat o}_{{2{N_g} + 1},{\rm{RSTO}}}}} \right]^T} $,~the RSTO is given as 
\begin{equation}
{\hat \tau _{\rm{R}}} = {\rm{index}}\left[ {\mathop {\arg \max }\limits_{1 \le i \le 2{N_g} + 1} \left( {{{\hat o}_{i,\rm{RSTO}}}} \right)} \right].
\end{equation}

\subsection{ELM-Based RCFO Estimation}
\begin{figure}[t]
	\centering
	\includegraphics[width=3.5in]{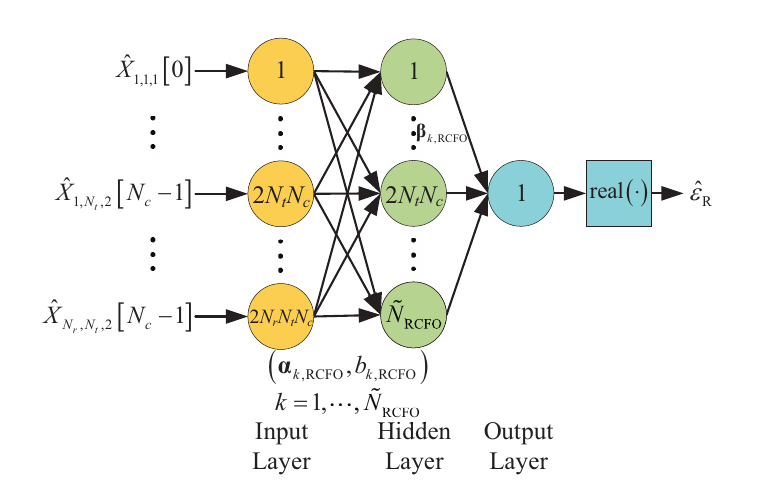}
	\caption{The structure of an ELM-based RCFO estimator.}
	\label{ELM-basedRCFOestimator}
\end{figure}

\begin{algorithm}[htb]
	\caption{The Training Algorithm for an ELM-based RCFO Estimator}
	\label{alg:2}
	\begin{algorithmic}
		\STATE We are given a training set ${\bf{N}}_{\rm{RCFO}} = \left\{ {\left( {{{{\bf{\tilde I}}}_n},{O_{n,{\rm{RCFO}}}}} \right)|n = 1, \cdots ,N_{\rm{RCFO}}} \right\}$, complex activation function $g_c\left( \cdot \right)$, and hidden neuron number $\tilde N_{\rm{RCFO}}$. ${\bf{\tilde I}}_n \in {{\mathbb{C}}^{2{N_r}{N_t}{N_c} \times 1}}$,~$ {O_{n,{\rm{RCFO}}}} $ is a real number and these two correspond to the input and desired output of the ELM-based RCFO estimator, respectively. Here, $ O_{n,{\rm{RCFO}}} $ denotes a given RCFO.
		\STATE {\bf{Steps 1, 2 and 3:}} Refer to {\bf{Algorithm \ref{alg:1}}}. 
	\end{algorithmic}
\end{algorithm}

As Fig. \ref{ELM-basedRCFOestimator} illustrates, the ELM-based RCFO estimator has $ 2N_rN_tN_c $ input neurons, $ \tilde N_{\rm{RCFO}} $ hidden neurons and only one output neuron.~Here, we use $ {\varepsilon _{\rm{R}}} $ to denote the RCFO, where $ {\varepsilon _{\rm{R}}}{\rm{ = }}\varepsilon  - \hat \varepsilon  $.

\subsubsection{Training Stage}
Before ELM can be deployed to estimate RCFO, it has to learn the prior knowledge from the training set. The training algorithm for an ELM-based RCFO estimator can be summarized as shown in \textbf{Algorithm \ref{alg:2}}.

Specifically,~the $ n $th input data of training set is given as
\begin{equation}
{{{\bf{\tilde I}}}_n} = {\mathop{\rm FFT}\nolimits} \left( {{\mathop{\rm Remove}\nolimits} {\mathop{\rm CP}\nolimits} \left( {{{\bf{I}}_n}} \right)} \right),
\end{equation}
where $ {{\bf{\tilde I}}_n} $~denotes the combination  vector of the preamble signal in the frequency domain by taking the FFT of the time domain received samples with corresponding RCFO,~$ {\varepsilon _{\rm{R}}} = {O_{n,{\rm{RCFO}}}} $. The time domain received samples with RCFO ${O_{n,{\rm{RCFO}}}}$ can be expressed as 
\begin{equation}
{{{\bf{I}}_n} = {{\left[ {{{{\bf{\tilde c}}}_1}^T, \cdots ,{{{\bf{\tilde c}}}_p}^T, \cdots ,{{{\bf{\tilde c}}}_{{N_t}}}^T} \right]}^T}\odot{{\left[ {{{{\bf{\tilde o}}}_1}^T, \cdots ,{{{\bf{\tilde o}}}_p}^T, \cdots ,{{{\bf{\tilde o}}}_{{N_t}}}^T} \right]}^T}},
\end{equation}
where
\begin{equation}
{{\bf{\tilde c}}_p} = {\rm{repmat}}\left( {{{\left[ {{\rm{CP}}_{{{\bf{c}}_{p,1}}}^T,{{\bf{c}}_{p,1}}^T,{\rm{CP}}_{{{\bf{c}}_{p,2}}}^T,{{\bf{c}}_{p,2}}^T} \right]}^T},{N_r},1} \right),
\end{equation}
\begin{equation}
{{{{\bf{\tilde o}}}_p} = {\rm{repmat}}\left( {{{\bf{o}}_p},{N_r},1} \right)},
\end{equation}
and
\begin{equation}
{{{\bf{o}}_p} = \left[ {\begin{array}{*{20}{c}}
		{{e^{2\pi j\left[ {1{\rm{ + 2}}\left( {p - 1} \right)\left( {{N_c} + {N_g}} \right)} \right]{O_{n,{\rm{RCFO}}}}/{N_c}}}}\\
		\vdots \\
		{{e^{2\pi j\left[ {2\left({N_c+{N_g}}\right) + 2 \left( {p - 1} \right)\left( {{N_c} + {N_g}} \right)} \right]{O_{n,{\rm{RCFO}}}}/{N_c}}}}
		\end{array}} \right]}.
\end{equation}
In {\bf{step 1}} of {\bf{Algorithm \ref{alg:2}}}, the generation of the input weight $ {\bm{\alpha}_{k,{\rm{RCFO}}}} $ and bias $ b_{k,{\rm{RCFO}}} $ are the same as {\bf{step 1}} in {\bf{Algorithm \ref{alg:1}}}. The output of the hidden layer can be given by

\begin{equation}
{{\bf{D}}_{{\rm{Training,RCFO}}}} = {g_c}\left( {{{\bm{\alpha }}^T_{\rm{RCFO}}}{\bm{{\rm \tilde I}}} + {\bf{b}}_{\rm{RCFO}}} \right),
\end{equation}
where $ {\bf{\tilde I}} = \left[ {\begin{array}{*{20}{c}}
	{{{\bf{\tilde I}}_1}}&{{{\bf{\tilde I}}_2}}& \cdots &{{{\bf{\tilde I}}_N}}
	\end{array}} \right] \in {{\mathbb{C}}^{2{N_r}{N_t}{N_c} \times N }}  $.

We would expect that the output of the ELM could be close to the target output $ \bf{O}_{\rm{RCFO}} $, so $ {\bm{\beta }}_{\rm{RCFO}}{{\bf{D}}_{{\rm{Training,RCFO}}}} = {\bf{O}}_{\rm{RCFO}} $. For the ELM-based RCFO estimator, $ N_{o,{\rm{RCFO}}}=1$. The LS solution is then given by
\begin{equation}
{\bm{\hat \beta }_{\rm{RCFO}}} = {\bf{O_{{\rm{RCFO}}}D}}_{{\rm{Training,RCFO}}}^\dag.
\end{equation}

\subsubsection{Prediction (Estimation) Stage}
Since the received preambles have been corrupted by the fading channel, channel estimation and equalization need to be carried out. Now that $ {\bf{c}}_{p,2} $ is known and fully occupies the subcarriers, the minimum mean-square error (MMSE) estimate of the frequency impulse response \cite{VanDeBeek1995} from the $ p $th TX antenna to the $ q $th RX antenna is given by
\begin{equation}
{\left[ {{{{\rm{\hat H}}}_{q,p}}\left[ 0 \right], \cdots ,{{{\rm{\hat H}}}_{q,p}}\left[ {{N_c} - 1} \right]} \right]^T} = {\bf{F}}{{\bf{Q}}_{{\rm{MMSE}}}}{{\bf{F}}^H}{\bf{c}}_{q,p,2}^H{{\bf{x}}_{q,p,2}},
\end{equation}
where ${\bf{Q}}_{\rm{MMSE}}$ can be shown to be
\begin{equation}
\begin{split}
{\bf{Q}}_{\rm{MMSE}}= {\bf{R}}_{\rm{gg}} \left[\left({\bf{F}}^H {\bf{c}}_{q,p,2}^H {\bf{c}}_{q,p,2} {\bf{F}} \right)^{-1}{\sigma^2_n}+{\bf{R}}_{\rm{gg}} \right]^{-1}\\ \times \left({\bf{F}}^H{\bf{c}}_{q,p,2}^H {\bf{c}}_{q,p,2} {\bf{F}} \right)^{-1}.
\end{split}
\end{equation}
${\bf{R}}_{\rm{gg}}$ is the auto-covariance matrix of $ {\left[ {{g_{q,p}}\left( 0 \right), \cdots ,{g_{q,p}}\left( {L - 1} \right)} \right]^T} $ for any $ \left( {q,p} \right) $. In other words, all the auto-covariance matrices of subchannels are the same.~$\sigma^2_n$ denotes the noise variance ${\rm{E}}\left\{\left|n^2_k\right|\right\}$. Here, we assume that the auto-covariance matrices of all the subchannels between the TX antennas and the RX antennas are the same, so the subscripts of ${\bf{R}}_{\rm{gg}}$ are omitted.  
Then, the equalized preamble can be given by
\begin{equation}
{\hat X_{q,p,i}}\left[ k \right] = {X_{q,p,i}}\left[ k \right]/{{\bf{\hat H}}_{q,p,i}}\left[ k \right].
\end{equation}
The calculation of the output of the hidden layer and $ {\bf{\hat x}} $ are same as Equations \eqref{Prediction} and \eqref{X}, respectively.
%The calculation of output of hidden layer can be calculated as
%\begin{equation}
%{{\bf{D}}_{{\rm{Prediction}}}} = {g_c}\left( {{{\bm{\alpha }}^T}{{\bf{\hat x}}} + {\bf{b}}} \right)
%\end{equation}
%where $ {\bf{\hat x}} $ is same as that in Equation \eqref{X}.
Finally, the RCFO is estimated as 
\begin{equation}
{\hat \varepsilon _{\rm{R}}} = {\mathop{\rm real}\nolimits} \left( {{\bm{\hat \beta }}_{\rm{RCFO}}{{\bf{D}}_{{\rm{Prediction,RCFO}}}}} \right).
\end{equation}

\subsection{Complexity Analysis}
We use the number of complex multiplications (CMs) to measure the computational complexity. For simplicity, the numbers of CMs for calculating the Moore-Penrose pseudoinverse of an $I \times O$ matrix is denoted as $C_{\rm{pinv}}\left(OI^2\right)$. We compare
the proposed complex ELM-based method with the DNN-based method. We assume that the input and output dimensions of the complex ELM-based method are $I$ and $O$, respectively. For a DNN-based method, we split a complex number into a real part and an imaginary part. Thus, the input and output dimensions of the DNN-based method are $2I$ and $2O$, respectively. In the DNN-based method, there are real-valued multiplications. When calculating the computational complexity, we consider that four real-valued multiplications are equivalent to one CM.

The machine learning-based method has two phases, i.e., the training phase and the prediction (estimation) phase, and we analyze the computational complexity for the two phases individually.

As for the training phase, the calculation of the output weights of complex ELM-based method requires $C_{\rm{pinv}}\left(N\tilde N^2\right)+N\left(\tilde N+O\right)$ CMs where $N$ is the number of training samples. The training complexity of the DNN-based method is difficult to derive using the number of CMs because it is trained iteratively with forward propagation (FP) and backpropagation (BP). Generally, the training complexity of the DNN is obviously higher than the complex ELM, and so the time consumption in training means it is difficult to satisfy the latency constraint in practical use cases.

The required numbers of CMs for the estimation phase are summarized in Table \ref{ComplexityAnalysis}, where $N_l$ and $n_l$ denote the number of hidden layers and the number of neurons at the $l$th hidden layer, respectively. $\tilde N$ is the number of hidden neurons in the complex ELM. As can be seen, the proposed method has a significantly lower computational complexity compared to the DNN-based method and only requires $IO{\tilde{N}}$ extra CMs compared to the traditional method.

%As for the prediction phase, the required number of CMs for the complex ELM and the DNN are $IO\tilde N$ and $ IO\sum\nolimits_{l = 1}^{{N_l}} {{n_l}{n_{l - 1}}} $, respectively. $N_l$ and $n_l$ denote the number of hidden layers and the number of neurons at the $l$th hidden layer, respectively. It can be seen that the prediction complexity of the DNN is higher than the complex ELM.%

Different from the stage of training, it should be noticed that the input of the ELM in the prediction stage is the preamble corrupted by RSTO, RCFO, multipath fading channel and thermal noise. Therefore, the only question is whether the proposed method can outperform the traditional method. The following subsection will provide exhaustive simulation results and comparisons.

\begin{table}[t]
	\centering
	\caption{COMPARISON OF COMPLEXITY IN THE ESTIMATION STAGE}
	\label{ComplexityAnalysis}
	\begin{tabular}{|l|l|}
		\hline
		\multicolumn{1}{|c|}{Methods} & \multicolumn{1}{c|}{Numbers of Multiplications} \\ \hline
		Traditional \cite{Schmidl1997,Zelst2004} & $N_tN_rN_c$ per metric calculation\\ \hline
		DNN-based & $N_tN_rN_c$ per metric calculation $+IO{\textstyle \sum_{l=1}^{N_l}} n_{l}n_{l-1}$ \\ \hline
		ELM-based & $N_tN_rN_c$ per metric calculation $+IO{\tilde{N}}$ \\ \hline
	\end{tabular}
\end{table}

\section{Simulation Results and Comparisons}
\label{S5}
In this section, the simulation setup is provided in detail and then, the performance of the proposed ELM-based RSTO and RCFO estimators is demonstrated.
\subsection{Simulation Setup}
\subsubsection{System Parameters and Channel Model}
For the numerical simulations, we set $ {N_c} = 64 $,~$ N_g=N_c/4 $ and sampling frequency $ f_s=4 \times {10^6} $. The wireless fading channel is modeled as an exponential model and quasistatic assumption is guaranteed during each OFDM symbol. For an exponentially decaying PDP, the root mean square (RMS) delay spread $ {\tau _{{\rm{RMS}}}}{\rm{ = 2}} \times {\rm{1}}{{\rm{0}}^{ - 6}}~{\rm{s}} $, and the coherence bandwidth $ {B_c} = 1/{\tau _{{\rm{RMS}}}} = {\rm{5}} \times {\rm{1}}{{\rm{0}}^5}~{\rm{Hz}} $. The PDP is given by
\begin{equation}
	{P_l} = \exp \left( {\frac{{ - {\rm{2}}\pi {B_c}{\tau _l}}}{{\sqrt 3 }}} \right)
\end{equation}
where the delay of $ l $th path is set as $ {\tau _l} = l{T_s} $ and $ L=8 $.

In addition, for a fair comparison, we keep the total transmitting power the same as in the single-input single-output (SISO) case. Therefore, the power per TX antenna is scaled down by a factor $ N_t $. 

\subsubsection{Training Sets, Hyperparameters and Activation Function}
The performance of a NN is usually sensitive to the training set, activation function and hyperparameters. Therefore, for an ELM-based RSTO and a RCFO estimators, the range and interval of the desired output in a training set should be chosen carefully. Generally, the power of a channel path with large delay is usually very low and therefore it may not significantly degrade the performance of timing synchronization. So we set the range of RSTO of training from $-N_g$ to $N_g$. Besides, the variance of the traditional CFO estimator decreases as $N_r$ increases. We also have observed that adjusting the range and interval of RCFO in a training set affects the performance of the ELM-based RCFO estimator. So we generate different training sets and use the method of grid search to choose the best one among them. The training sets for different MIMO systems are summarized in Table \ref{RCFOTrainingSets}. The number of hidden neurons $\tilde{N}$ is selected by the same method. Moreover, $\tilde{N}$ must be larger than or equal to the size of a training set $N$ so that the capacity of an ELM is high enough to learn the training set \cite{huang2003learning}. We choose ${g_c}\left( z \right)={\rm{arcsinh}}\left(z\right)=\int_0^zdt/\left[\left(1+t^2\right)^{1/2}\right]$, where $z \in \mathbb{C}$ as the complex activation function, because arcsinh performs very well in certain learning environments due to its elegant symmetric and ``squashing'' magnitude responses \cite{Kim2003}.

\begin{table}[t]
	\centering
	\caption{Parameters of Training Sets and $\tilde N$ for Different MIMO Systems}
	\label{RCFOTrainingSets}
	\begin{tabular}{|l|c|c|c|}
		\hline
		\multicolumn{1}{|c|}{MIMO System} & Range of RSTO & Interval of RSTO & $\tilde N_{\rm{RSTO}}$ \\ \hline
		$2 \times 2$                               & $\left[ { - {N_g},{N_g}} \right]$        & $1$ & $2^{14}$           \\ \hline
		\multicolumn{1}{|c|}{MIMO System} & Range of RCFO & Interval of RCFO & $\tilde N_{\rm{RCFO}}$ \\ \hline
		$1 \times 1$                               & $\left[ { - {\rm{0}}{\rm{.0025,0}}{\rm{.0025}}} \right]$        & $ 5.0 \times {{10}^{ - 6}} $ & $2^{11}$            \\ \hline
		$2 \times 2$                               & $\left[ { - {\rm{0}}{\rm{.0025,0}}{\rm{.0025}}} \right]$        & $ 2.5 \times {{10}^{ - 6}} $ & $2^{14}$            \\ \hline
		$3 \times 3$ (Fading)              & $ \left[ { - {\rm{0}}{\rm{.0030,0}}{\rm{.0030}}} \right] $        & $ 5.0 \times {{10}^{ - 6}} $ & $2^{17}$             \\ \hline
		$3 \times 3$ (AWGN)                & $ \left[ { - {\rm{0}}{\rm{.05,0}}{\rm{.05}}} \right] $          & $ 1.0 \times {{10}^{ - 4}} $ & $2^{17}$             \\ \hline
		$4 \times 4$ (Fading)                & $ \left[ { - {\rm{0}}{\rm{.0010,0}}{\rm{.0010}}} \right] $          & $ 1.0 \times {{10}^{ - 5}} $ & $2^{14}$             \\ \hline
		$4 \times 4$ (AWGN)                & $ \left[ { - {\rm{0}}{\rm{.05,0}}{\rm{.05}}} \right] $          & $ 5.0 \times {{10}^{ - 4}} $ & $2^{14}$             \\ \hline
	\end{tabular}
\end{table}
\subsection{Performance of the ELM-based RSTO Estimator}
As it is instructive to observe the mean bias error (MBE) as well as the MSE of an estimator (where $ {\rm{MSE}} = {\rm{variance}} + {\left( {{\rm{bias}}} \right)^2} $), we will examine both the bias and the MSE of the proposed estimator and compare this with the traditional estimator. 
\begin{figure}[t]
	\centering
	\includegraphics[width=3.4in]{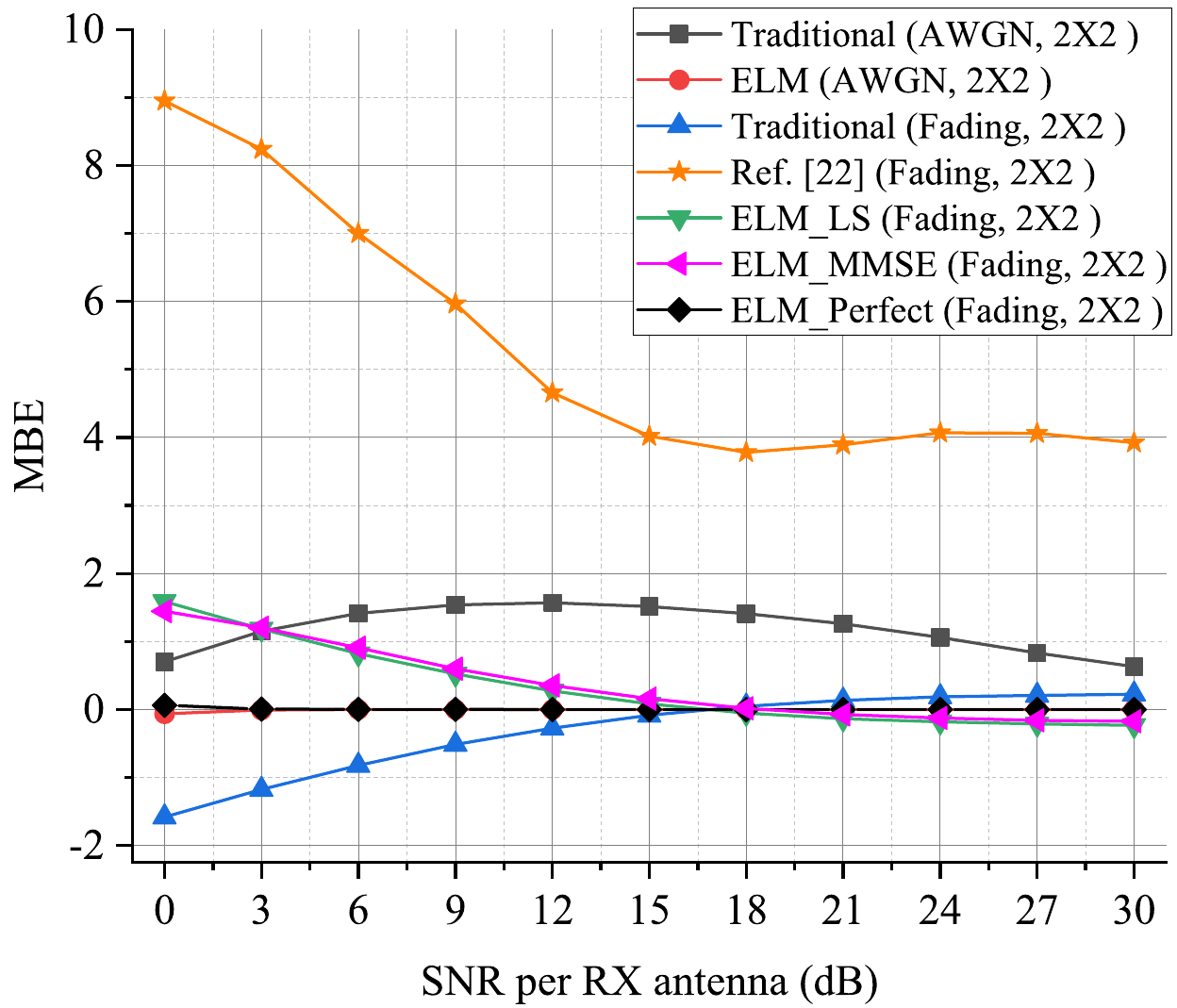}
	\caption{MBE performance comparison between the traditional STO estimator, ELM-based estimator in \cite{Qing2020} and the proposed ELM-based STO estimator for a $ 2\times 2 $ system with AWGN and a multipath fading channel.}
	\label{RSTOMean2X2}
\end{figure}

\begin{figure}[t]
	\centering
	\includegraphics[width=3.4in]{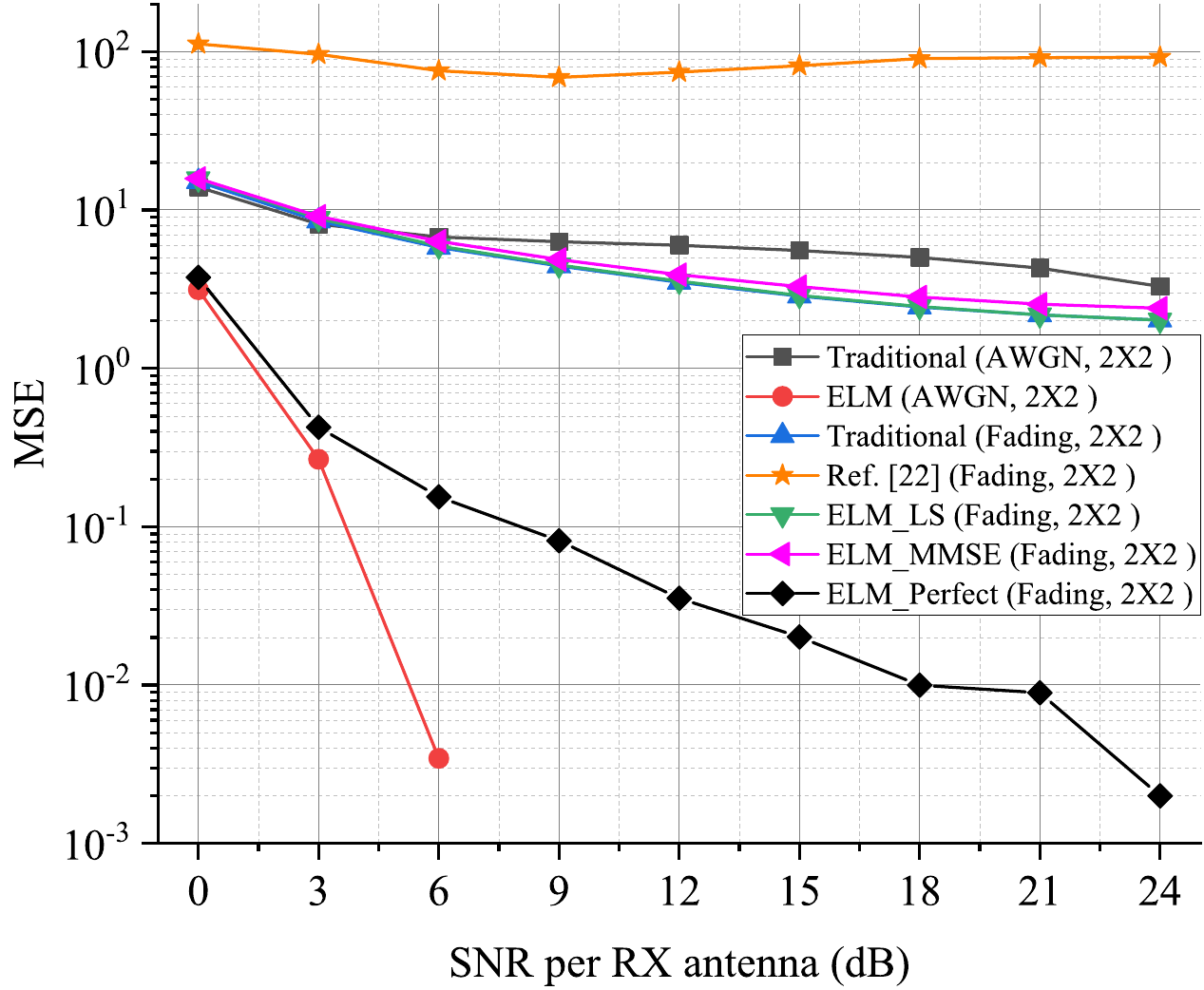}
	\caption{MSE performance comparison between the traditional STO estimator, ELM-based estimator in \cite{Qing2020} and the proposed ELM-based STO estimator for a $ 2 \times 2 $ system with AWGN and a multipath fading channel.}
	\label{RSTOMSE2X2}
\end{figure}

In Fig. \ref{RSTOMean2X2}, the MBE of the STO estimation is illustrated as a function of the average SNR per receive antenna. The results from Monte Carlo simulations averaged over $ {3\times10^5} $ channel realizations are shown. In the cases of both an AWGN channel and a frequency-selective fading channel with perfect CSI information, it can be seen that the proposed ELM-based estimator has a much smaller bias than the traditional estimator. The MBE of ELM-based estimator approaches zero when $\rm{SNR}\geq3~dB$. In the cases of a frequency-selective fading channel with LS and MMSE channel estimation, the proposed ELM-based STO estimator does not show a gain in terms of MBE. This is because the imperfect channel estimation leads to the error of timing metric. In addition, the MBE of a recent ELM-based estimator proposed in \cite{Qing2020} is obviously higher than other estimators even at a relatively high SNR because the method of the generation of training data in \cite{Qing2020} relies on a given channel model and therefore makes it impractical. 

In Fig. \ref{RSTOMSE2X2}, the MSE of the STO estimation is displayed as a function of the average SNR per receive antenna. It can be seen that the proposed ELM-based estimator has a significantly smaller MSE than the traditional estimator and the ELM-based estimator in \cite{Qing2020} in the cases of both an AWGN channel and a frequency-selective fading channel with perfect CSI information. Even if the proposed ELM-based estimator acquires the imperfect CSI just by using a LS or an MMSE channel estimate, the MSE performance of the proposed ELM-based STO estimator is much better than the ELM-based estimator in \cite{Qing2020}. It shows that the proposed method can learn the RSTO based on training data and then predict it very well.

\subsection{Performance of the ELM-based RCFO Estimator }
For the estimation of CFO, the performance of the estimator based on the traditional method and its Cramér-Rao lower bound (CRLB) will also be studied and used as a benchmark. Note that the CRLB  is equal to the variance of traditional CFO estimator \cite{Zelst2004}
\begin{equation}
	{\mathop{\rm var}} \left( {\hat \varepsilon  - \varepsilon } \right) = \frac{1}{{{\pi ^2}{N_t}{N_r}V\rho }} 
	\label{CRLB}
\end{equation}
where $ V $ is the length of identical halves in the first part of the preamble and ~$ \rho  = \left( {P/{N_t}} \right)\sigma _n^2 $ denotes the SNR per receive antenna when the preamble is transmitting and $ P $ is the total transmit power.\footnote{\eqref{CRLB} also can be written as ${\rm{var}}\left( {\hat \varepsilon  - \varepsilon } \right) = \frac{1}{{{\pi ^2}{N_r}V\rho '}}$ where the SNR per receive antenna is defined as $ \rho ' = P\sigma _n^2 $.} Note that \eqref{CRLB} is approximately accurate under the condition of small errors $ \left( {\hat \varepsilon  - \varepsilon } \right) $ and high SNR and it is derived in the appendix. We assume that timing synchronization is perfect~($ \hat{\tau}=\tau $) ~when the performances of both the traditional and the ELM-based CFO estimators are evaluated.
\subsubsection{MSE Performance}
\begin{figure}[t]
	\centering
	\includegraphics[width=3.4in]{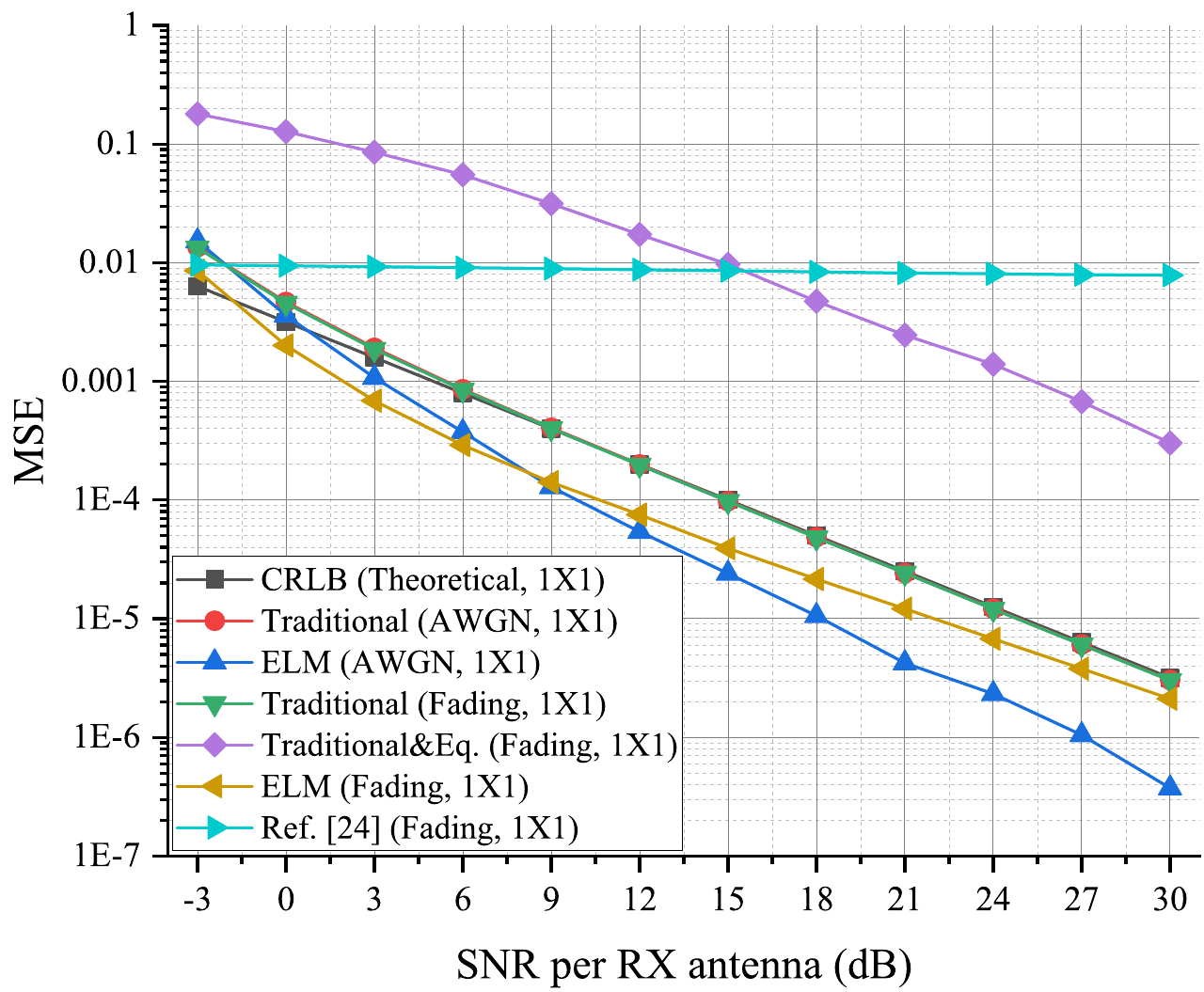}
	\caption{MSE performance comparison between the traditional CFO estimator, DNN-based estimator in \cite{Ninkovic2020} and the proposed ELM-based CFO estimator for an $ 1 \times 1 $ system from theory and simulations with AWGN and a multipath fading channel.}
	\label{RCFOMSE1X1}
\end{figure}
%($ \varepsilon {\rm{ = }} - 0.05 $, $ {O_{n,{\rm{RCFO}}}} \in \left\{ {\left. {5.0 \times {{10}^{ - 6}}k} \right|k =  - 500, \cdots ,500} \right\} $ and $ \tilde N_{\rm{RCFO}} = {2^{11}} $)%

\begin{figure}[htb]
	\centering
	\includegraphics[width=3.4in]{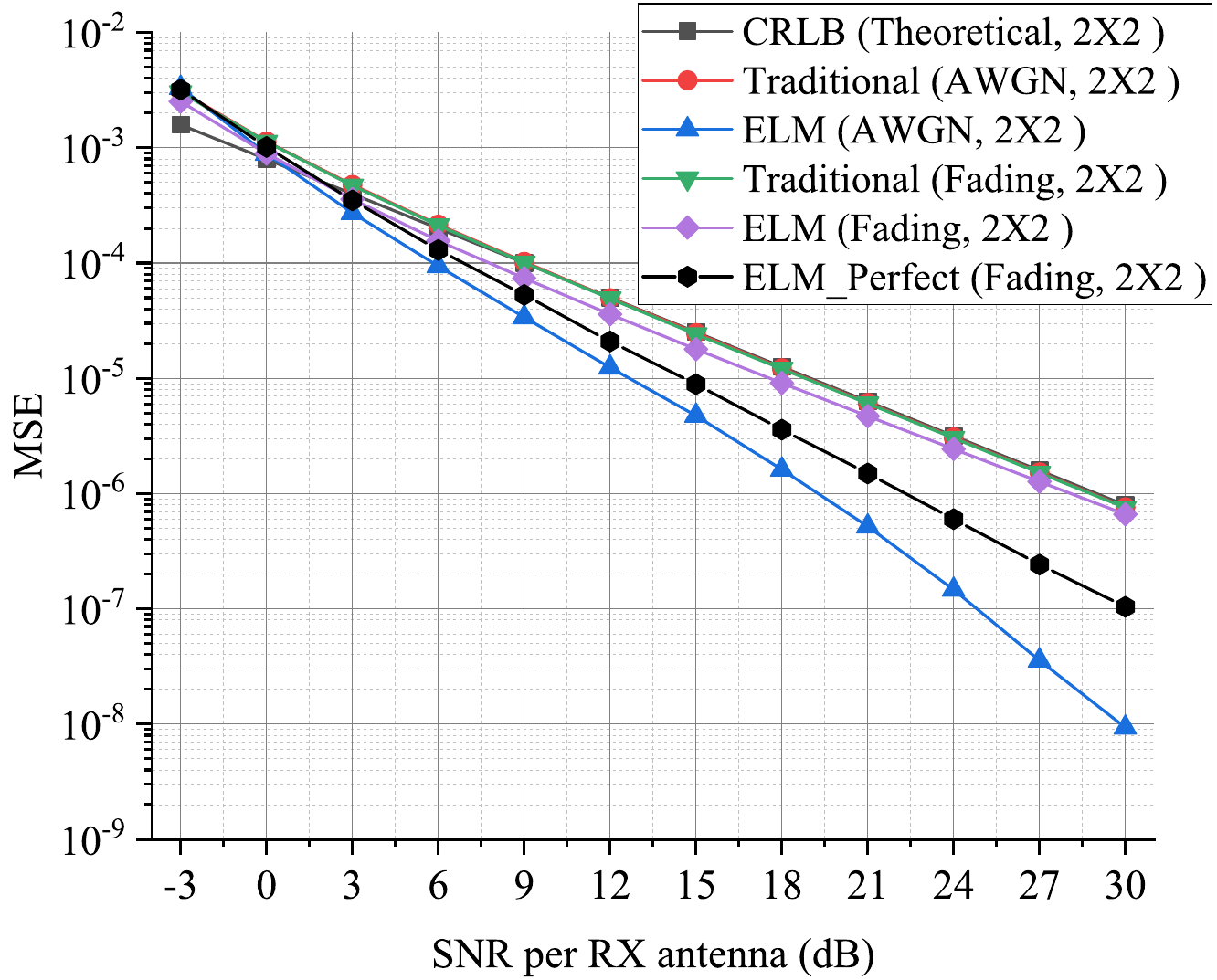}
	\caption{MSE performance comparison between the traditional CFO estimator and the proposed ELM-based CFO estimators for a $ 2 \times 2 $ system from theory and simulations with AWGN and a multipath fading channel.}
	\label{RCFOMSE2X2}
\end{figure}
%($ \varepsilon {\rm{ = }} - 0.05 $, $ {O_{n,{\rm{RCFO}}}} \in \left\{ {\left. {2.5 \times {{10}^{ - 6}}k} \right|k =  - 1000, \cdots ,1000} \right\} $ and $ \tilde N_{\rm{RCFO}} = {2^{14}} $)%
In Fig. \ref{RCFOMSE1X1}, the MSE of the CFO estimation is demonstrated as a function of the average SNR per receive antenna. The theoretical value from (\ref{CRLB}) is shown together with results from Monte Carlo simulations averaged over $ {10^5} $ channel realizations. As seen from Fig. \ref{RCFOMSE1X1}, the MSE curve of the traditional method almost perfectly overlaps that of the CRLB. The theoretical value is a good estimate of the MSE for high SNR values but underestimates the MSE compared with simulation results for low SNR. Note that the CRLB expresses a lower bound on the variance of unbiased estimators of a deterministic parameter. A biased approach can result in both a variance and a MSE that are below the unbiased CRLB. Specifically, in the case of AWGN, when $ \rm{SNR}=-3~\rm{dB} $, ELM obtains a slightly larger MSE value compared with simulation results for the traditional method. Fortunately, we can see that the  performance improvement between ELM and the traditional method increases with SNR, and when $ \rm{SNR}=21~\rm{dB} $ ELM achieves a SNR gain of about 9~dB over the traditional method at a MSE value of $ 4.22\times10^{-6} $. In the case of a frequency-selective fading channel, the largest SNR gain over a traditional method, about 4.5~dB, is achieved when $ \rm{SNR}=6~\rm{dB} $. This kind of the gain becomes insignificant when $ {\rm{SNR > 27~dB}} $.

We also use the DNN-based model and training method proposed in \cite{Ninkovic2020} to estimate RCFO. It can be seen that its MSE performance is significantly inferior to that of other methods and it does not have obvious change as SNR increases. This is because the channel impulse used for the generation of training set is different from the channel impulse of each channel realization, and therefore it leads to poor performance.

To provide further insights on the proposed method and prove that the gain of ELM is not just a result of channel estimation and equalization, the curve of ``Traditional\&Eq.'' shows the MSE performance of the traditional method with channel estimation and equalization. Specifically, the method ``Traditional\&Eq.'' performs traditional CFO estimation twice. The first CFO estimation uses the traditional method. Then, channel estimation and equalization are performed by using the frequency corrected preamble signal. Finally, the traditional CFO estimation method is performed again to estimate RCFO by using the frequency corrected and equalized preamble signal. It can be seen that its performance seriously degrades, which means that channel estimation and equalization cannot enhance the performance of the traditional CFO estimator. This can be explained by the fact that the ELM can reuse the preamble exhaustively and also learn the mapping relationships between RCFOs and their corresponding preambles with RCFOs.

Fig. \ref{RCFOMSE2X2} presents the MSE curves of the traditional and proposed ELM-based CFO estimators for a $ 2 \times 2 $ system. Similar to the observations in Fig. \ref{RCFOMSE1X1}, in the case of an AWGN channel, the MSE performance of the ELM still outperforms that of the traditional method. Using the ELM, when $ \rm{SNR}=18~\rm{dB} $, about 9~dB SNR gain over traditional method is achieved at $ \rm{MSE}=2.16\times10^{-6} $. In the case of a frequency-selective fading channel, by comparing Fig. \ref{RCFOMSE2X2} with Fig. \ref{RCFOMSE1X1}, we find that the gains of ELM over the traditional method for a $ 2 \times 2 $ system (about 1.5~dB) are lower than that for $ 1\times1 $ system (about 1.5-4.5~dB). We conjecture that this is because the accuracy of channel estimation limits the CFO estimation performance of the ELM.

In order to prove this conjecture, we also perform the simulation for the  ELM estimator with perfect channel state information (CSI). In Fig. \ref{RCFOMSE2X2}, the curve ``ELM\_Perfect'' illustrates the performance of the ELM with perfect CSI. It can be seen that, under the condition of knowing perfect CSI, the MSE performance of the ELM under the condition of fading channel is closer to that under the condition of AWGN channel compared to that of ELM without perfect CSI. The gain from perfect CSI increases with the increase of SNR. So we can conjecture that the performance of the ELM-based scheme is sensitive to the accuracy of the CSI.

\begin{figure}[t]
	\centering
	\includegraphics[width=3.4in]{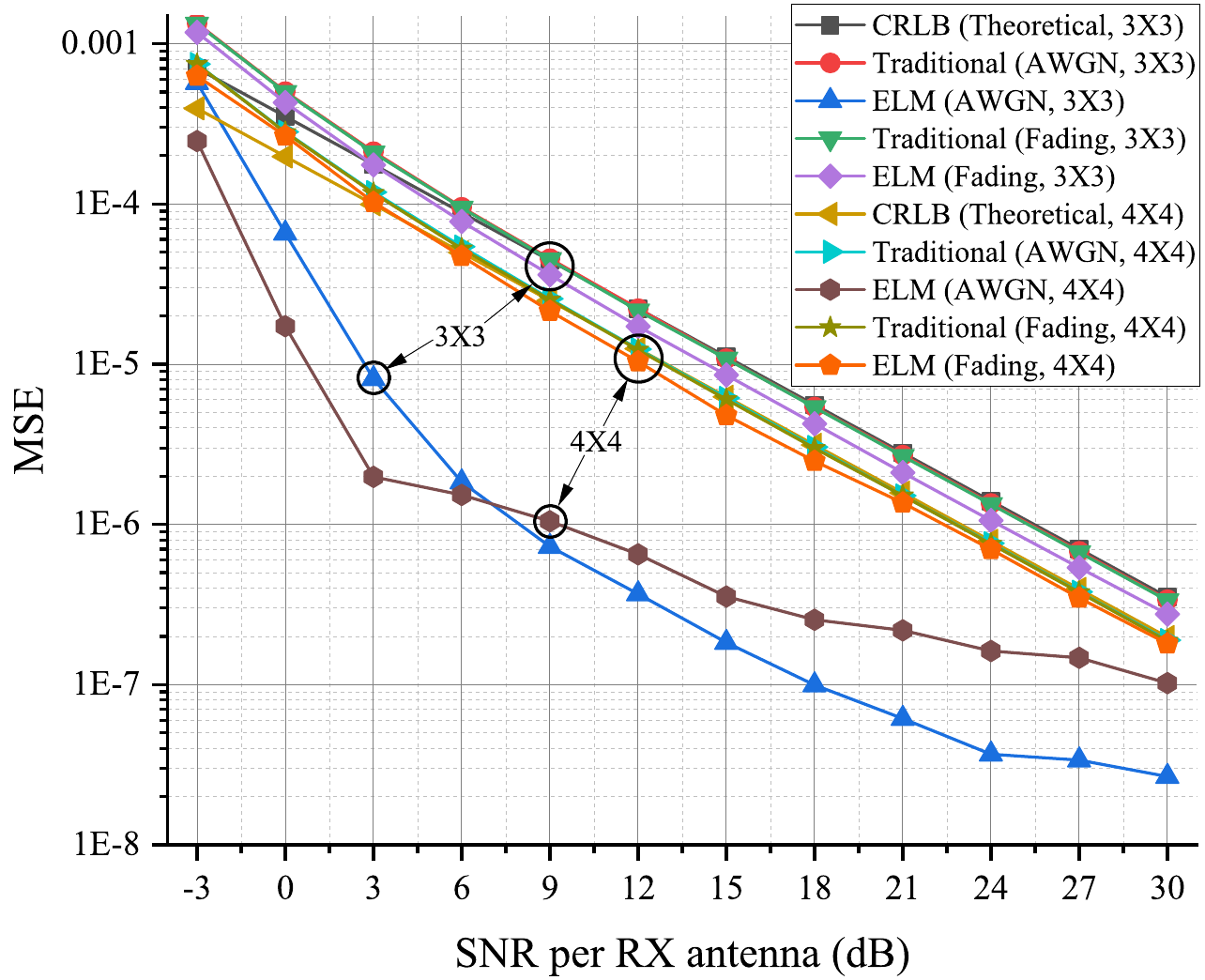}
	\caption{MSE performance comparison between the traditional CFO estimator and the proposed ELM-based CFO estimator for a $ 3 \times 3 $ and $ 4 \times 4 $ systems from theory and simulations with AWGN and a multipath fading channel.}
	\label{RCFOMSE3X3&4X4}
\end{figure}
%($ \varepsilon {\rm{ = }} - 0.05 $, $ \tilde N_{\rm{RCFO}} = {2^{17}} $ and AWGN: $ {O_{n,{\rm{RCFO}}}} \in \left\{ {\left. {1.0 \times {{10}^{ - 4}}k} \right|k =  - 500, \cdots ,500} \right\} $;~Fading: $ {O_{n,{\rm{RCFO}}}} \in \left\{ {\left. {5.0 \times {{10}^{ - 6}}k} \right|k =  - 600, \cdots ,600} \right\} $)%

Fig. \ref{RCFOMSE3X3&4X4} presents the MSE curves of the traditional and the proposed ELM-based CFO estimators for a $ 3\times3 $ and $ 4\times4 $ MIMO systems. Note that for the $ 3\times3 $ and $ 4\times4 $ systems, the ELM models are trained by different training sets separately in order to achieve the best performance under AWGN and fading channel conditions. In the case of an AWGN channel for the $ 3 \times 3 $ system, when $ {\rm{SNR = }} - {\rm{3~dB}} $, the MSE performance of the ELM is slightly better than the CRLB but its MSE decreases rapidly with an increase of SNR. When $ {\rm{SNR = 12~dB}} $, about 18~dB SNR gain over the traditional method is achieved by the ELM-based method. For the $ 4 \times 4 $ system, ELM-based method achieves its highest gain (about 16.5 dB) over the traditional method. By comparing Fig. \ref{RCFOMSE3X3&4X4} with Fig. \ref{RCFOMSE2X2} and Fig. \ref{RCFOMSE1X1}, the gain of the ELM-based method over the traditional method at low SNRs increases with the increase in the number of RX antennas. At high SNRs, the gains are not as high as the highest gains. This is because the MSE performance of the ELM relates to the number of the receive antennas, and the gain of ELM is from both noise suppression and mining information exhaustively from preambles. In the case of a frequency-selective fading channel for the both $ 3\times3 $ and $ 4\times4 $ systems, the ELM can obtain about 1.5~dB gain of MSE when ${\rm{SNR}} \ge ~{\rm{3dB}}$, which is similar with the case for a $ 2\times2 $ system.

\subsubsection{Robustness Analysis}
\begin{figure}[t]
	\centering
	\includegraphics[width=3.4in]{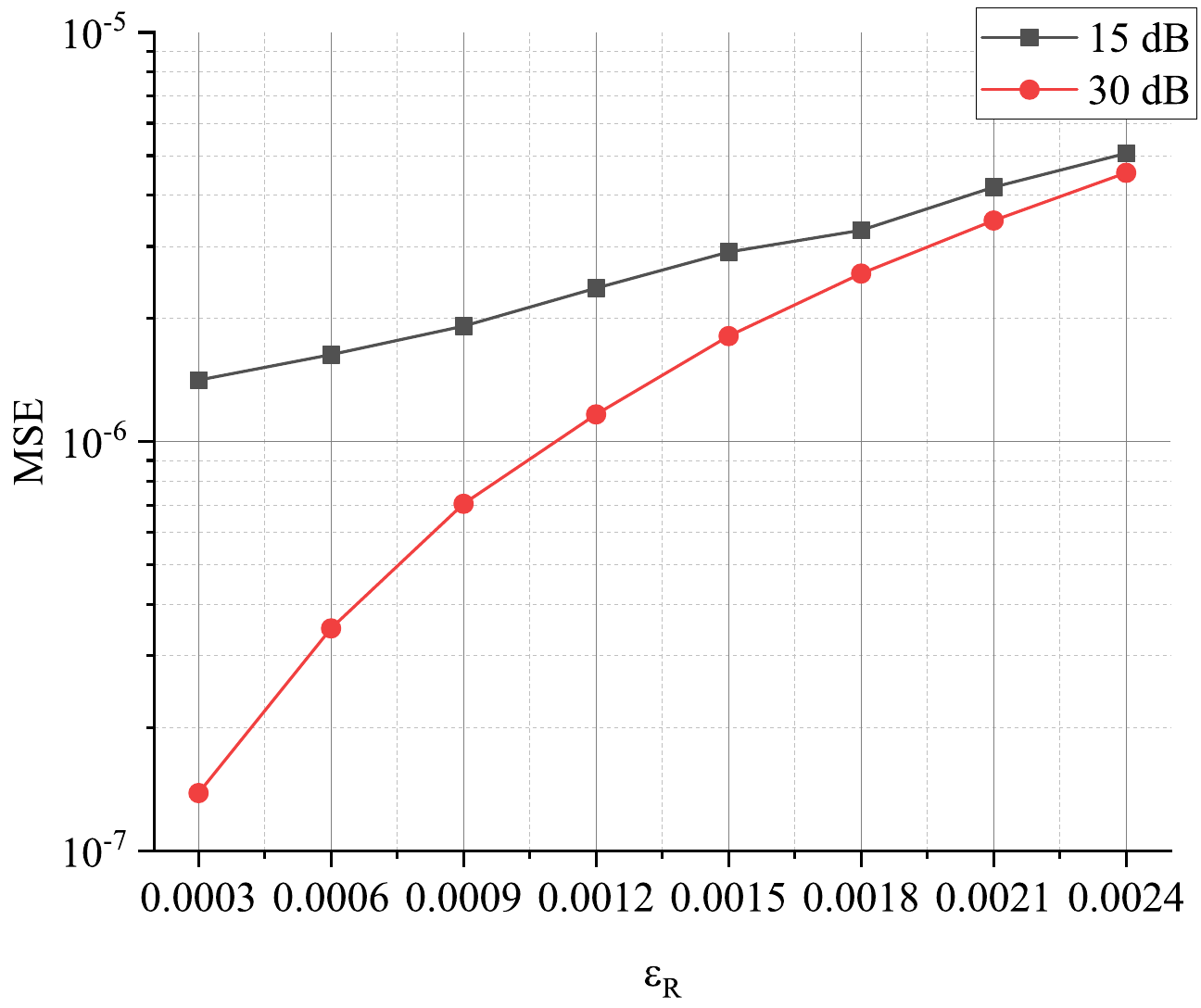}
	\caption{MSE versus ${\varepsilon _{\rm{R}}}$ curves of the proposed ELM-based CFO estimator for a $ 2 \times 2 $ system.}
	\label{RobustnessRCFO}
\end{figure}
%($ {O_{n,{\rm{RCFO}}}} \in \left\{ {\left. {2.5 \times {{10}^{ - 6}}k} \right|k =  - 1000, \cdots ,1000} \right\} $ and $ \tilde N_{\rm{RCFO}} = {2^{14}} $)%

\begin{figure}[t]
	\centering
	\includegraphics[width=3.4in]{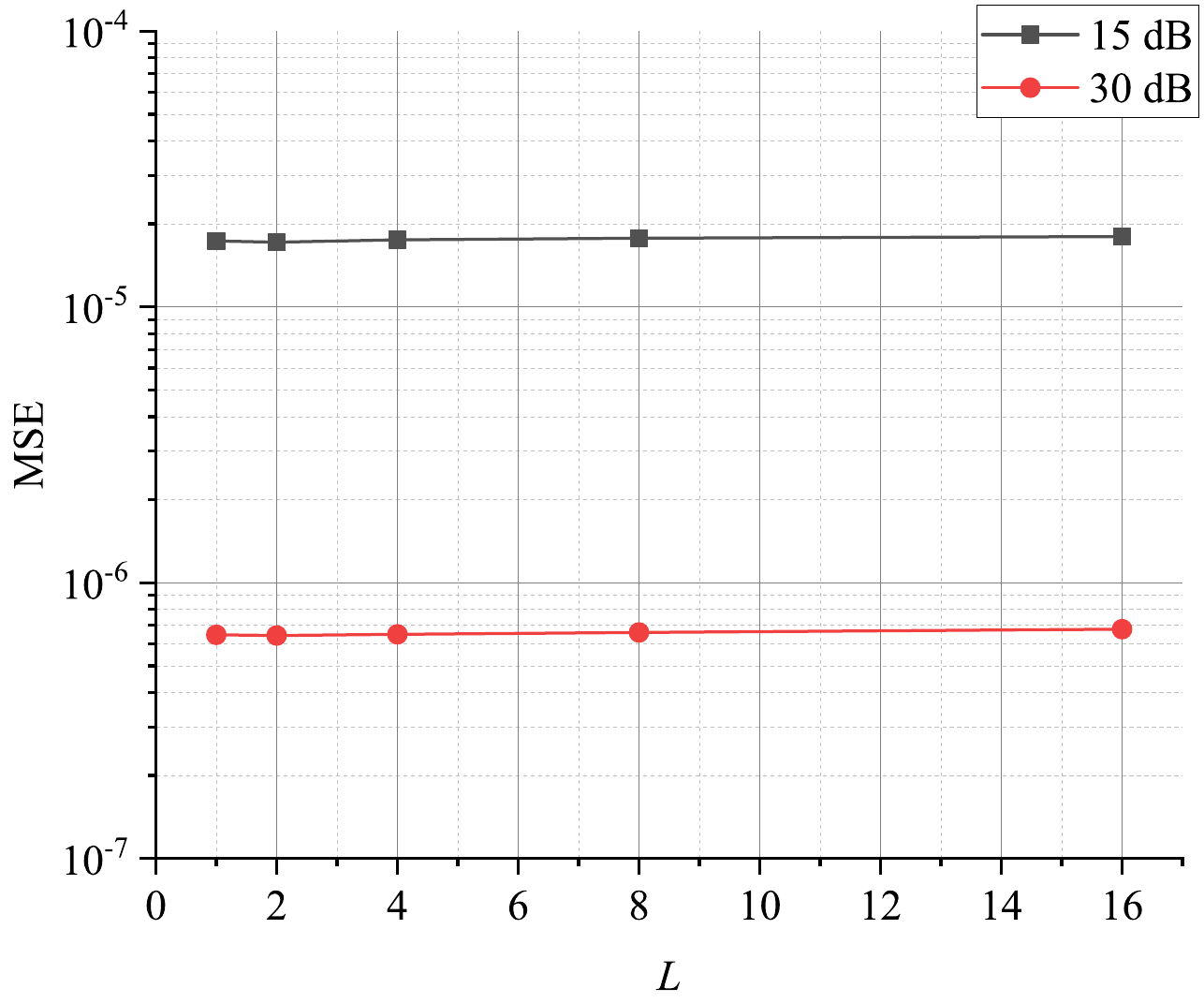}
	\caption{MSE versus the number of paths of a fading channel $ \left( L \right) $ of the proposed ELM-based CFO estimator for a $ 2 \times 2 $ system.}
	\label{RobustnessChannel}
\end{figure}
%($ {O_{n,{\rm{RCFO}}}} \in \left\{ {\left. {2.5 \times {{10}^{ - 6}}k} \right|k =  - 1000, \cdots ,1000} \right\} $ and $ \tilde N_{\rm{RCFO}} = {2^{14}} $)%
In this part, we analyze the robustness of the proposed ELM scheme.	Fig. \ref{RobustnessRCFO} shows the MSE of ELM under various RCFOs when $ {\rm{SNR = 15~dB}} $~and $ {\rm{30~dB}} $. It can be seen that the MSE increases with an increase of RCFO. By comparing Fig. \ref{RobustnessRCFO} with Fig. \ref{RCFOMSE2X2}, when $ {\rm{SNR = 30~dB}} $ and $  {\varepsilon _{\rm{R}}}\ge 0.0012 $, the MSE of the ELM is higher than that in Fig. \ref{RCFOMSE2X2} ($ {\rm{MSE}}> {10^{ - 6}} $). However, when $ {\rm{SNR = 15~dB}} $ and $  {\varepsilon _{\rm{R}}}\le 0.0024 $, the MSE of ELM is lower than that in Fig. \ref{RCFOMSE2X2} ($ {\rm{MSE}}< {10^{ - 5}} $). This can be explained by the fact that the performance advantage and robustness of the ELM-based method are more significant in medium SNR.

Fig. \ref{RobustnessChannel} shows the MSE of the ELM under a channel with different number of channel paths when $ {\rm{SNR = 15~dB}} $~and $ {\rm{30~dB}} $. It can be seen that, with increase of $ L $, the MSE of the ELM is almost unchanged, which means that the proposed method is robust enough to handle frequency-selective fading channels with different numbers of paths. The simulation results demonstrate the robustness of the proposed method to the RCFO and propagation environments.

\subsubsection{Generalization Error Analysis}
\begin{figure}[t]
	\centering
	\includegraphics[width=3.4in]{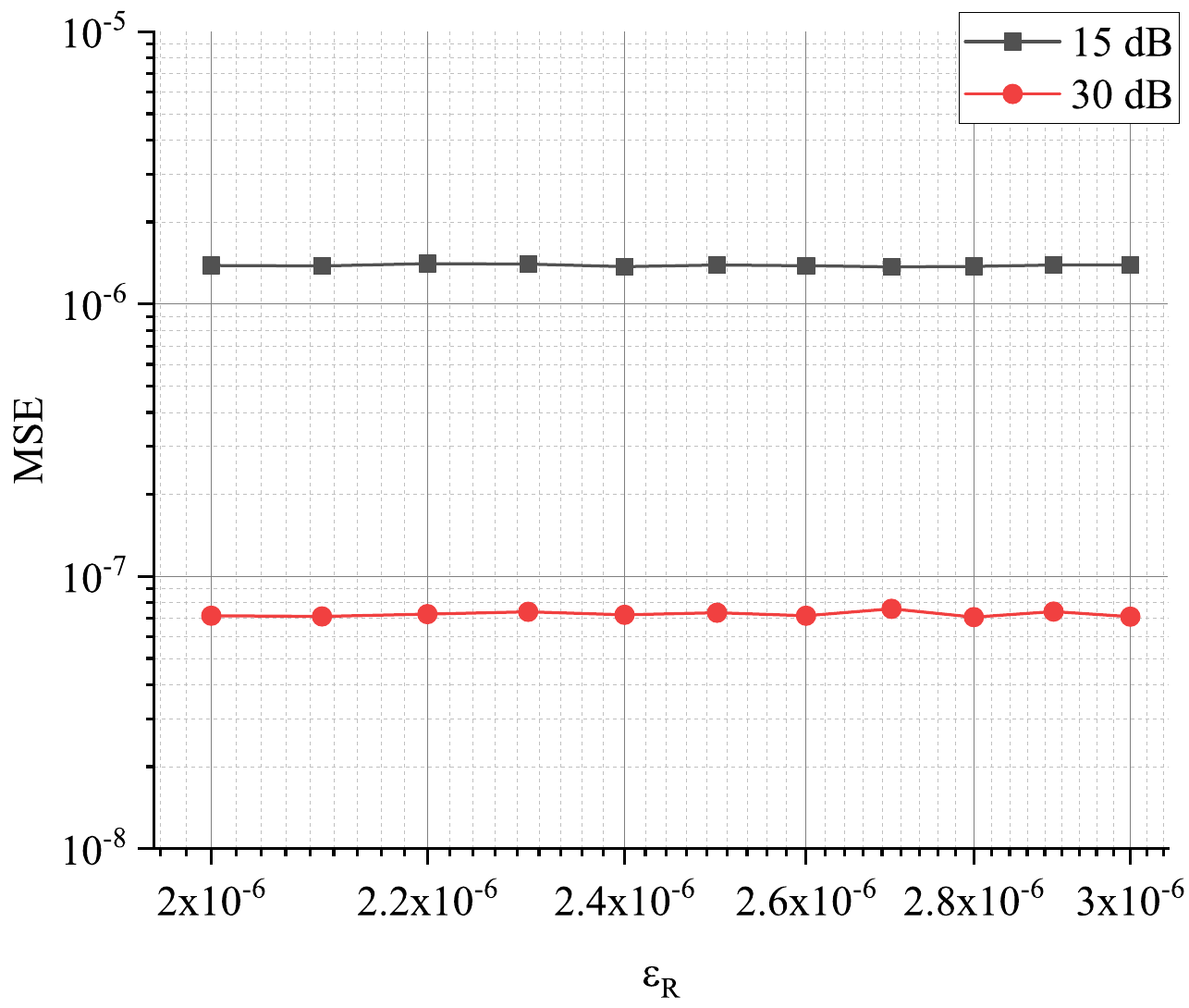}
	\caption{MSE versus the RCFO not belonging to a training set of the proposed ELM-based CFO estimator for a $ 2 \times 2 $ system.}
	\label{GeneralizationRCFO}
\end{figure}
%($ {O_{n,{\rm{RCFO}}}} \in \left\{ {\left. {2.5 \times {{10}^{ - 6}}k} \right|k =  - 1000, \cdots ,1000} \right\} $ and $ \tilde N_{\rm{RCFO}} = {2^{14}} $)%

Generalization is a term used to describe the ability of a model to react to new data. That is, after being trained on a training set, can a model ``digest'' new data and make accurate predictions? In this paper, we also use MSE to evaluate the generalization ability of the ELM. Specifically, an RCFO not belonging to the training set is used to verify the generalization of a trained ELM-based RCFO estimator. The generalization ability of ELM can be analyzed according to Fig. \ref{GeneralizationRCFO}, where we have used RCFOs that do not belong to the training set $ {O_{n,\rm{RCFO}}} \in \left\{ {\left. {2.5 \times {{10}^{ - 6}}k} \right|k =  - 1000, \cdots ,1000} \right\} $. By comparing Fig. \ref{GeneralizationRCFO} with Fig. \ref{RCFOMSE2X2}, the performance of the ELM-based RCFO estimator does not change significantly when it handles those unfamiliar RCFOs. It can be concluded that the ELM-based RCFO estimator shows excellent generalization when it processes an RCFO not belonging to the training set.

\section{Conclusions and future work}
\label{S6}
In this paper, we have proposed an ELM-based fine timing and frequency synchronization scheme in order to improve the performance of existing estimators. The proposed scheme does not require any additional preamble and the training processes can be carried out fully offline without any prior information about the channels. Simulation results have shown that the proposed ELM-based synchronization scheme outperforms or achieves comparable performance in terms of MSE with existing traditional and learning-based synchronization algorithms. In addition, the proposed scheme shows robustness under various channels with different parameters and a generalization ability towards any RCFO outside the training set.

The simulation results have shown that the performance of the proposed ELM-based scheme relates to the accuracy of the CSI. Besides, it should be noticed that channel equalization can neutralize the effects of both STO and the fading channel. Therefore, this makes it difficult to obtain the received preamble signal affected by just the STO alone. In other words, the accurate CSI is still indispensable for the deployment of machine learning in communications systems. Therefore, incorporating ELM into transmitter design, synchronization and channel estimation and equalization jointly within system design is a promising future research direction.

%\subsection{Subsection Heading Here}
%Subsection text here.

%\IEEEpubidadjcol

%\subsubsection{Subsubsection Heading Here}
%Subsubsection text here.

% or


\appendices
\section{Variance of the CFO Estimation for a MIMO System Under AWGN Channel~(See \eqref{CRLB})}
We use the method in \cite{Moose1994} and \cite{schenk2008rf} to derive the variance of the CFO estimate for a MIMO system. According to \eqref{Estimation CFO}, for a given $ \varepsilon $, subtract the corresponding phase, $ \pi\varepsilon $, from each product to obtain the tangent of the phase error
\begin{equation}
\tan \left[ {\pi \left( {\hat \varepsilon  - \varepsilon } \right)} \right] = \frac{{\sum\limits_{p = 1}^{{N_t}} {\sum\limits_{q = 1}^{{N_r}} {\sum\limits_{i = {d_p} - \left( {V - 1} \right)}^{{d_p}} {{\rm{Im}}\left[ {r_q^*\left( {i - V} \right){r_q}\left( i \right){e^{ - \pi j\varepsilon }}} \right]} } } }}{{\sum\limits_{p = 1}^{{N_t}} {\sum\limits_{q = 1}^{{N_r}} {\sum\limits_{i = {d_p} - \left( {V - 1} \right)}^{{d_p}} {{\rm{Re}}\left[ {r_q^*\left( {i - V} \right){r_q}\left( i \right){e^{ - \pi j\varepsilon }}} \right]} } } }}
\label{Appendix1}
\end{equation}
where ${d_p} = d - \left( {{N_t} - p} \right){N_{{\rm{train}}}}$,~$ V $ denotes the length of identical halves in the first part of the preamble and $ {r_q}\left( i \right) = {{\tilde r}_q}\left( i \right) + {n_q}\left( i \right) $. $ {n_q}\left( i \right) $ denotes the time domain noise of $ i $th sample of the received signal on the $ q $th antenna. For$ \left| {\hat \varepsilon  - \varepsilon } \right| \ll 1/\pi  $, the tangent can be approximated by its argument so that
\begin{equation}
\resizebox{.95\hsize}{!}{$ \hat \varepsilon  - \varepsilon  \approx \frac{{\sum\limits_{p = 1}^{{N_t}} {\sum\limits_{q = 1}^{{N_r}} {\sum\limits_{i = {d_p} - \left( {V - 1} \right)}^{{d_p}} {{\rm{Im}}\left\{ {\left[ {{{\tilde r}_q}\left( {i - V} \right) + {n_q}\left( i \right){e^{ - \pi j\varepsilon }}} \right]\left[ {\tilde r_q^*\left( i \right) + {n_q}\left( i \right){e^{ - \pi j\varepsilon }}} \right]} \right\}} } } }}{{\pi \sum\limits_{p = 1}^{{N_t}} {\sum\limits_{q = 1}^{{N_r}} {\sum\limits_{i = {d_p} - \left( {V - 1} \right)}^{{d_p}} {{\rm{Re}}\left\{ {\left[ {{{\tilde r}_q}\left( {i - V} \right) + {n_q}\left( i \right){e^{ - \pi j\varepsilon }}} \right]\left[ {\tilde r_q^*\left( i \right) + {n_q}\left( i \right){e^{ - \pi j\varepsilon }}} \right]} \right\}} } } }}. $}
\label{Appendix2}
\end{equation}
According to the method in \cite{Moose1994}, at high SNR, a condition compatible with successful communications signalling means that \eqref{Appendix2} may be approximated by 
\begin{equation}
\resizebox{.95\hsize}{!}{$ \hat \varepsilon  - \varepsilon  \approx \frac{{\left\{ {\sum\limits_{p = 1}^{{N_t}} {\sum\limits_{q = 1}^{{N_r}} {\sum\limits_{i = {d_p} - \left( {V - 1} \right)}^{{d_p}} {{\rm{Im}}\left[ {{n_q}\left( i \right)\tilde r_q^*\left( {i - V} \right){e^{ - \pi j\varepsilon }} + {{\tilde r}_q}\left( {i - V} \right)n_q^*\left( {i - V} \right)} \right]} } } } \right\}}}{{\left\{ {\pi \sum\limits_{p = 1}^{{N_t}} {\sum\limits_{q = 1}^{{N_r}} {\sum\limits_{i = {d_p} - \left( {V - 1} \right)}^{{d_p}} {{{\left| {{{\tilde r}_q}\left( i \right)} \right|}^2}} } } } \right\}}}. $}
\label{Appendix3}
\end{equation}
It is easy to show that
\begin{equation}
{\rm E}\left[ {\left. {\hat \varepsilon  - \varepsilon } \right|\varepsilon ,\left\{ {{{\tilde r}_q}} \right\}} \right] = 0.
\label{Appendix4}
\end{equation}
Therefore, for small errors, the estimate is conditionally unbiased. Then, the conditional variance of the estimate is easily determined for \eqref{Appendix3} as
\begin{equation}
{\rm{Var}}\left[ {\left. {\hat \varepsilon } \right|\varepsilon ,\left\{ {{{\tilde r}_q}} \right\}} \right] = \frac{1}{{{\pi ^2}{N_t}{N_r}V\rho }}.
\label{Appendix5}
\end{equation}
Finally,~note that in this paper, $ \rho  = \sigma _{_{\tilde r,q}}^2/\sigma _n^2 = \left( {P/{N_t}} \right)\sigma _n^2 $ denotes the SNR per receive antenna when the preamble is transmitting and $ P $ is the total transmit power.
%\section{}

\section*{Acknowledgement}
The authors would like to acknowledge the help received from Longguang Wang. In addition, Jun Liu gratefully acknowledges the financial support received from the China Scholarship Council (CSC) and the School of Electronic and Electrical Engineering, University of Leeds, UK. He also wants to thank, in particular, the inspiration and care received from Yanling (Julia) Zhu during the period of this COVID-19 pandemic.

\ifCLASSOPTIONcaptionsoff
  \newpage
\fi



%\bibliographystyle{IEEEtran}

%\begin{thebibliography}{1}
%\end{thebibliography}

\bibliographystyle{IEEEtran}  %这是你要使用的格式,比如要投IEEE,就写IEEEtran
\bibliography{IEEEabrv,References/MyCollection}%这个是加载你的bib,你可以理解从文献数据库中加载要引用的文献

% your own custom macro containing the udegraphics\incl command to make things
\begin{IEEEbiography}[{\includegraphics[width=1in,height=1.25in,clip,keepaspectratio]{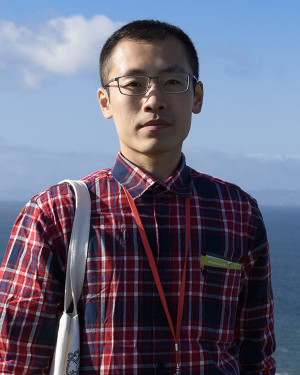}}]{Jun Liu}
received the B.S. degree in optical information science and technology from the South China University of Technology (SCUT), Guangzhou, China, in 2015, and the M.E. degree in communications and information engineering from the National University of Defense Technology (NUDT), Changsha, China, in 2017, where he is currently pursuing the Ph.D. degree with the Department of Cognitive Communications. 

He was a visiting Ph.D. student with the University of Leeds from 2019 to 2020. His current research interests include machine learning with a focus on shallow neural networks applications, signal processing for broadband wireless communication systems, multiple antenna techniques, and wireless channel modeling.
\end{IEEEbiography}

\begin{IEEEbiography}[{\includegraphics[width=1in,height=1.25in,clip,keepaspectratio]{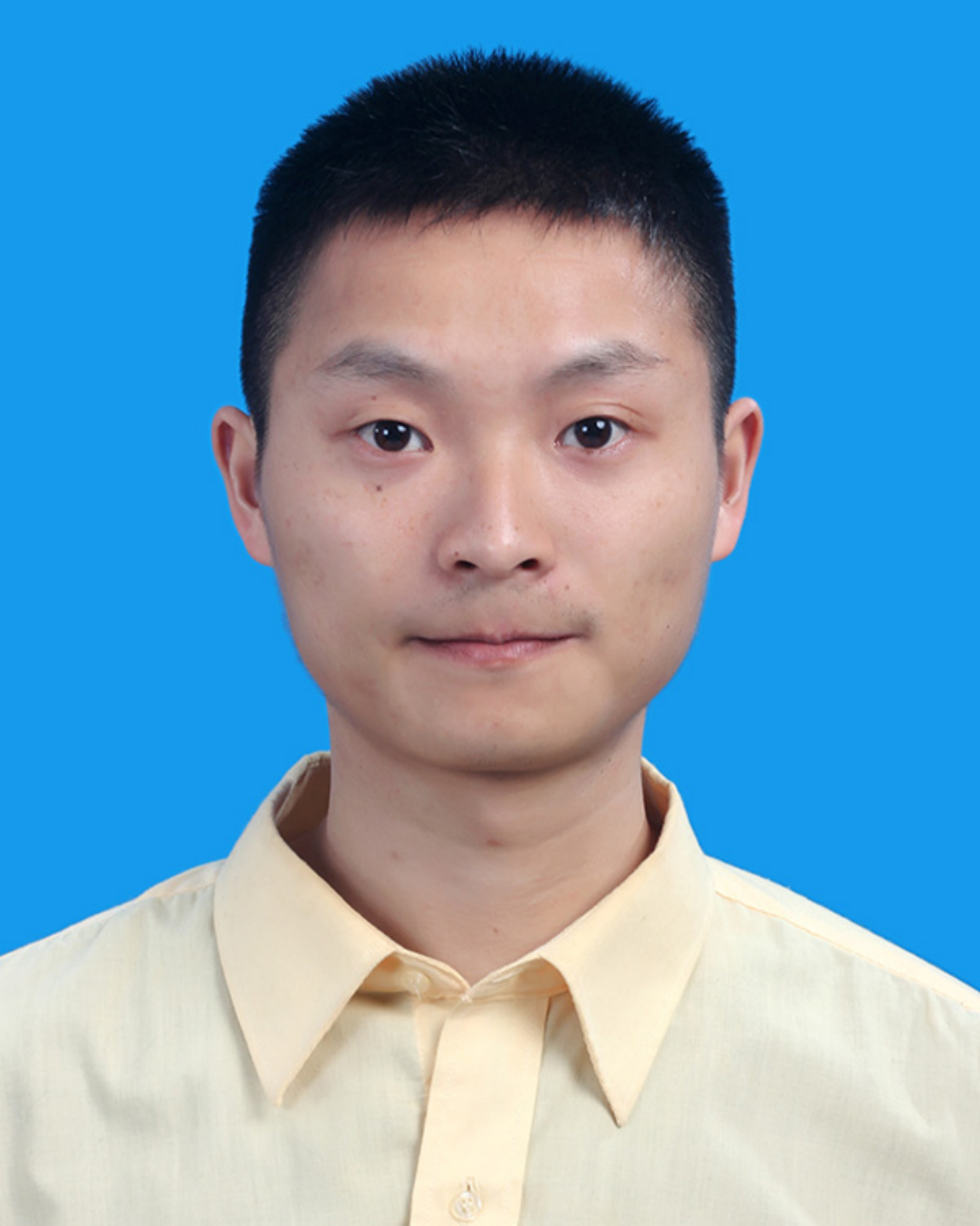}}]{Kai Mei}
received the master’s degree from the National University of Defense Technology, in 2017, where he is currently pursuing the Ph.D. degree. His research interests include synchronization and channel estimation in OFDM systems and MIMO-OFDM systems, and machine learning applications in wireless communications.
\end{IEEEbiography}

\begin{IEEEbiography}[{\includegraphics[width=1in,height=1.25in,clip,keepaspectratio]{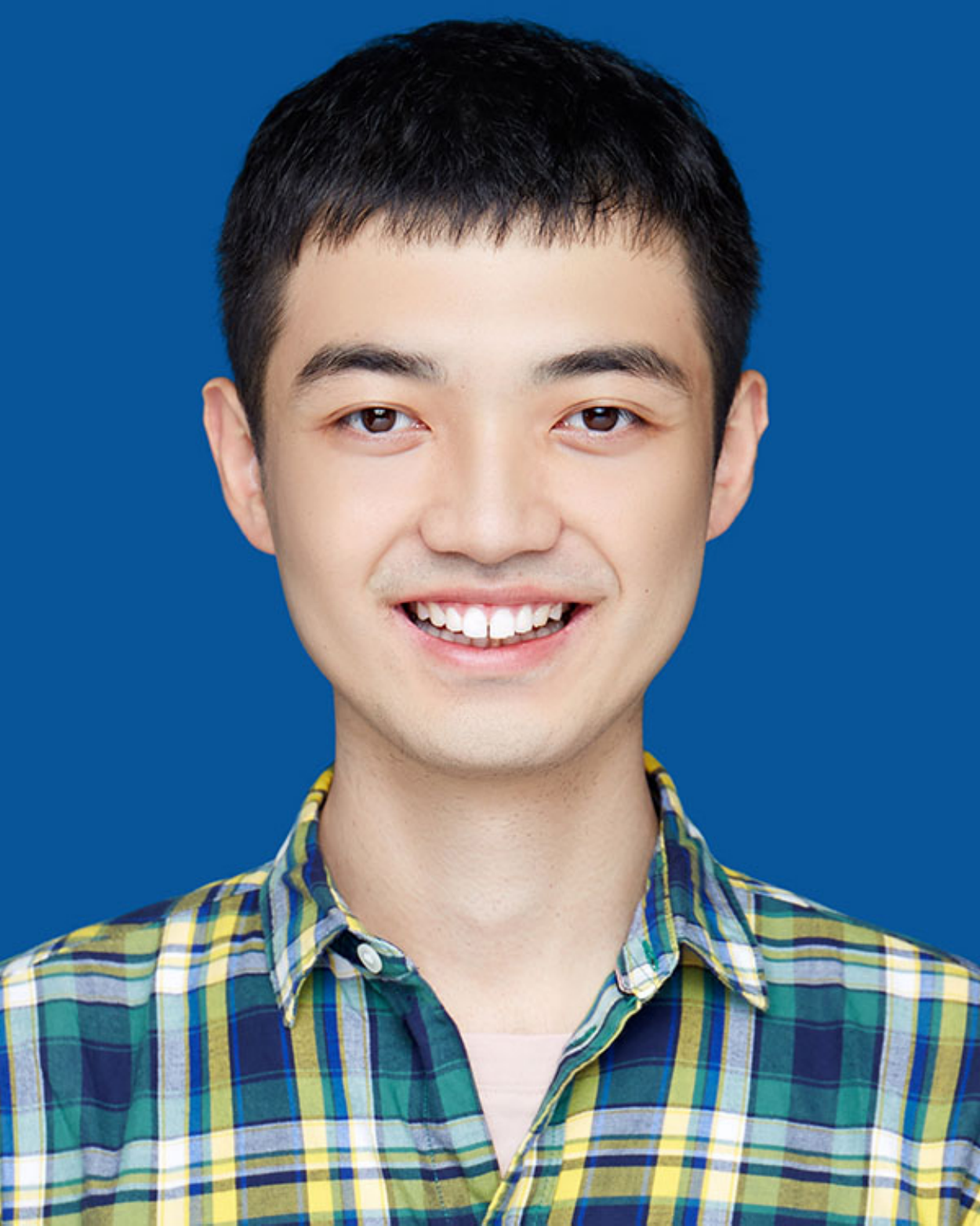}}]{Xiaochen Zhang}
is currently a graduate student at College of Electronic Science and Engineering from National University of Defense Technology (NUDT), Changsha, China. He received the B.S. degree from NUDT in 2018. His research interests include resource allocation, multi-access edge computing, machine learning and channel modeling.
\end{IEEEbiography}

\begin{IEEEbiography}[{\includegraphics[width=1in,height=1.25in,clip,keepaspectratio]{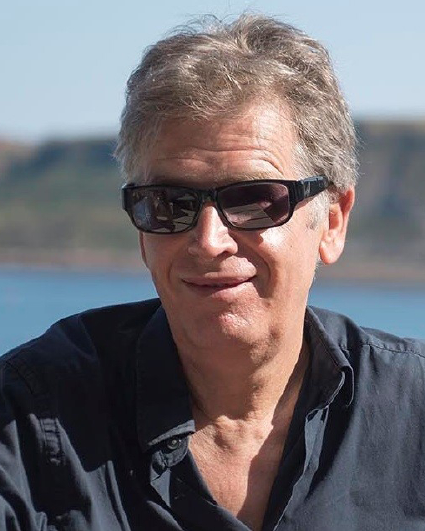}}]{Des McLernon}
(Member, IEEE) received his B.Sc and MSc degrees from the Queen’s University of Belfast, N. Ireland. After working on radar systems research with Ferranti Ltd in Edinburgh, Scotland, he then joined Imperial College, University of London, UK, where he took his PhD in signal processing. His research interests are broadly within the domain of signal processing for wireless communications in which field he has published around 340 journal and conference papers. Finally, in what spare time remains, he plays jazz piano in restaurants and bars and was recently runner-up in the 2018 “Leeds Pub Piano” competition.
\end{IEEEbiography}

\begin{IEEEbiography}[{\includegraphics[width=1in,height=1.25in,clip,keepaspectratio]{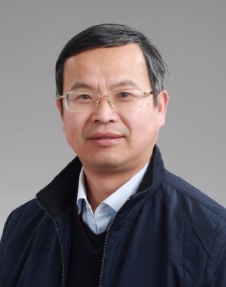}}]{Dongtang Ma}
(SM’13) received the B.S. degree in applied physics and the M.S. and Ph.D. degrees in information and communication engineering from the National University of Defense Technology (NUDT), Changsha, China, in 1990, 1997, and 2004, respectively. From 2004 to 2009, he was an Associate Professor with the College of Electronic Science and Engineering, NUDT. Since 2009, he is a professor with the department of cognitive communication, School of Electronic Science and Engineering, NUDT. From Aug. 2012 to Feb. 2013, he was a visiting professor at University of Surrey, UK. His research interests include wireless communication and networks, physical layer security, intelligent communication and network. He has published more than 150 journal and conference papers. He is one of the Executive Directors of Hunan Electronic Institute. He severed as the TPC member of PIMRC from 2012 to 2020.
\end{IEEEbiography}

\begin{IEEEbiography}[{\includegraphics[width=1in,height=1.25in,clip,keepaspectratio]{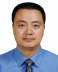}}]{Jibo Wei}
(Member, IEEE) received the B.S. and M.S. degrees from the National University of Defense Technology (NUDT), Changsha, China, in 1989 and 1992, respectively, and the Ph.D. degree from Southeast University, Nanjing, China, in 1998, all in electronic engineering. He is currently the Director and a Professor of the Department of Communication Engineering, NUDT. His research interests include wireless network protocol and signal processing in communications, more specially, the areas of MIMO, multicarrier transmission, cooperative communication, and cognitive network. He is a member of the IEEE Communication Society and also a member of the IEEE VTS. He also works as one of the editors of the Journal on Communications and is a Senior Member of the China Institute of Communications and Electronics.
\end{IEEEbiography}

\begin{IEEEbiography}[{\includegraphics[width=1in,height=1.25in,clip,keepaspectratio]{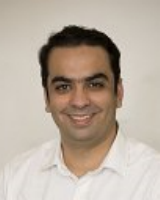}}]{Syed Ali Raza Zaidi}
	(Senior Member, IEEE) is currently a University Academic Fellow (Assistant Professor) in the broad area of Communication and Sensing for RAS. He was awarded J. W. and F. W. Carter Prize, was also awarded with COST IC0902, EPSRC, DAAD and Royal Academy of Engineering grants. He has published more than 100 technical papers in various top-tier IEEE Journals and conferences. His research interests include design and implementation of communication protocols for wireless networking specifically in the area of M2M.
\end{IEEEbiography}

%\newpage



%\vfill


\end{document}